\documentclass[10pt,twocolumn,aps,prx,superscriptaddress,longbibliography,eprint]{revtex4-2}

\usepackage{graphicx}
\usepackage{amsmath}
\usepackage{amssymb}
\usepackage{float}
\usepackage{braket}   
\usepackage{enumitem}
\usepackage[dvipsnames]{xcolor}
\usepackage[colorlinks=true, urlcolor=blue, linkcolor=blue,citecolor=blue, hypertexnames=false]{hyperref}
\usepackage[detect-weight=true,separate-uncertainty=true]{siunitx}
\usepackage{comment}
\usepackage[normalem]{ulem}

\usepackage{cleveref}
\crefname{equation}{Eq.}{Eqs.}
\Crefname{equation}{Equation}{Equations}
\crefname{figure}{Fig.}{Figs.}
\Crefname{figure}{Figure}{Figures}
\crefname{section}{Sec.}{Sects.}
\Crefname{section}{Section}{Sections}
\crefname{table}{Table}{Tables}
\crefname{appendix}{Appendix}{Apps.}
\Crefname{appendix}{Appendix}{Apps.}

\newcommand{\h}[1]{\hat{#1}}
\newcommand{\ha}{\hat{a}}
\newcommand{\had}{\hat{a}^\dagger}
\newcommand{\hb}{\hat{b}}
\newcommand{\hbd}{\hat{b}^\dagger}

\newcommand{\hH}{\hat{H}}
\newcommand{\sx}{\hat\sigma_x}
\newcommand{\sz}{\hat\sigma_z}

\newcommand{\hvt}{\hat{\varphi}_{t}}
\newcommand{\hvr}{\hat{\varphi}_{r}}
\newcommand{\zpfr}{\varphi_{{\rm rzpf}}}
\newcommand{\zpft}{\varphi_{{\rm tzpf}}}

\begin{document}

\title{Robustness of longitudinal transmon readout to ionization}

\author{Alex A. Chapple}
\email{alex.arimoto.chapple@usherbrooke.ca}
\affiliation{Institut Quantique and D\'epartement de Physique, Universit\'e de Sherbrooke, Sherbrooke J1K 2R1 QC, Canada}
\author{Alexander McDonald}
\affiliation{Institut Quantique and D\'epartement de Physique, Universit\'e de Sherbrooke, Sherbrooke J1K 2R1 QC, Canada}
\author{Manuel H. Muñoz-Arias}
\altaffiliation{Peresent address: Quantum Algorithms and Applications Collaboratory, Sandia National Laboratories, Livermore, CA 94550, USA}
\affiliation{Institut Quantique and D\'epartement de Physique, Universit\'e de Sherbrooke, Sherbrooke J1K 2R1 QC, Canada}
\author{Mathieu Lachapelle}
\altaffiliation{Present address: Canadian Centre for Climate Modelling and Analysis, Environment and Climate Change Canada, Victoria, British Columbia, Canada.}
\affiliation{Institut Quantique and D\'epartement de Physique, Universit\'e de Sherbrooke, Sherbrooke J1K 2R1 QC, Canada}
\author{Alexandre Blais}
\affiliation{Institut Quantique and D\'epartement de Physique, Universit\'e de Sherbrooke, Sherbrooke J1K 2R1 QC, Canada}
\affiliation{CIFAR, Toronto, ON M5G 1M1, Canada}

\date{\today}

\begin{abstract}

Multi-photon processes deteriorate the quantum non-demolition (QND) character of the dispersive readout in circuit QED, causing readout to lag behind single and two-qubit gates, in both speed and fidelity. Alternative methods such as the longitudinal readout have been proposed, however, it is unknown to what extent multi-photon processes hinder this approach. Here we investigate the QND character of the longitudinal readout of the transmon qubit. We show that the deleterious effects that arise due to multi-photon transitions can be heavily suppressed with detuning, owing to the fact that the longitudinal interaction strength is independent of the transmon-resonator detuning. We consider the effect of circuit disorder, the selection rules that act on the transmon, as well as the description of longitudinal readout in the classical limit of the transmon to show qualitatively that longitudinal readout is robust. We show that fast, high-fidelity QND readout of transmon qubits is possible with longitudinal coupling.
\end{abstract}

\maketitle

\section{Introduction} \label{sec:intro}

Fast and high-fidelity quantum non-demolition (QND) qubit readout is essential for fault-tolerant quantum computers~\cite{Terhal2015}. The standard method for qubit measurement in circuit quantum electrodynamics (cQED) is the dispersive readout, in which  a resonator coupled to the qubit undergoes a qubit-state-dependent frequency shift~\cite{RMP}. By probing this frequency shift with a measurement pulse, the qubit state can be inferred. Optimization of the dispersive readout has resulted in readout fidelity exceeding $99.9\%$ in an integration time of less than $60$ ns \cite{spring_readout,swiadek2023enhancing,Walter_disp_readout,Sunada_intrinsic}. Despite this tremendous progress, readout lags behind in performance compared to other qubit operations such as single- and two-qubit gates~\cite{Li_SQG, li2024_TQG, Sung_TQG, Somoroff_fluxonium_SQG}. An important difficulty in improving the dispersive readout is that, at short times, the response of the resonator only weakly depends on the qubit state, making it difficult to shorten the measurement time. Even more problematic is the breakdown of the QND character of this readout observed at small to moderate measurement photon number filling the resonator~\cite{MIST_1, MIST_2, Walter_disp_readout,Minev2019,Lescanne2019,Thorbeck2024}.  

Roughly speaking, this can be traced back to the transverse nature of the qubit-resonator coupling $\hat H_x =  g_x(\ha^\dagger+\ha)\sx$ at the origin of this readout. In the dispersive regime, where the qubit-resonator frequency detuning $\Delta$ is large compared to the coupling $g_x$, the transverse interaction can be pertubatively reduced to $\hat{H}_\chi = \chi_x \ha^\dagger\ha\sz$, with $\chi_x = g_x^2/\Delta$ the dispersive coupling. While this approximate dispersive Hamiltonian is QND, its parent Hamiltonian $\hH_x$ is not, leading to qubit state transitions~~\cite{Dynamics_of_transmon_ionzation, dumas2024unified, MIST_1, MIST_2,Verney2019,Lescanne2019,Xiao2023}.
More precisely, the root cause of the above issue is multiphoton resonances resulting in measurement-induced transition to highly excited transmon states~\cite{MIST_1}, something which has also been referred to as transmon ionization~\cite{dumas2024unified,Dynamics_of_transmon_ionzation,Reminiscence_chaos}. 
Moreover, when combined with the probe tone, the dispersive interaction $\hat{H}_\chi$ leads to qubit-state conditional trajectories of the resonator in phase space that are initially nearly parallel. 
This results in little information gain at short times, and thus the need for long integration times. In addition, to maximize the resonator pointer state separation in steady state, and therefore the information gain at long times, the dispersive shift must be set to $2\chi_x = \kappa$, with $\kappa$ the resonator decay rate~\cite{Gambetta2008}. Although this can easily be done in practice, this is yet another constraint that needs to be satisfied in experiments. 

The longitudinal readout is an alternative qubit measurement scheme  introduced to alleviate these issues~\cite{Didier_longitudinal, Mathieu_thesis}. In contrast to the dispersive readout, it exploits a longitudinal qubit-resonator coupling, $\hH_z = g_z(\ha^\dagger+\ha)\sz$~\cite{Richer_longitudinal,Kerman2013,Billangeon2015, potts:2024, salunkhe:2025}. Because the interaction $\hH_z$ is proportional to $\sz$, it is ideally QND. In addition, $\hH_z$ generates the optimal displacements of the two qubit-state-dependent resonator states that are 180 degrees out of phase, allowing for short measurement times~\cite{Munoz-Arias_cloaking_readout}. The absence of dressing between the qubit and resonator state under the longitudinal interaction has the added advantage of avoiding Purcell decay~\cite{Kerman2013,Billangeon2015}. Furthermore, the longitudinal interaction relaxes the matching condition $2\chi_x=\kappa$ of the dispersive readout. One approach to obtain an effective longitudinal interaction is to strongly drive a resonator dispersively coupled to a qubit, the strong drive displacing the cavity field $\ha\rightarrow\ha+\alpha$~\cite{RMP,Blais2007}. For large amplitudes of the cavity field $\alpha$, this leads to a dominantly longitudinal interaction with $g_z = \alpha\chi$. However, if the dispersive interaction originates from a transversal interaction, this approach is not free from the limitations mentioned above. 

The advantages of longitudinal readout over dispersive readout were derived under the assumption of an ideal two-level system~\cite{Didier_longitudinal}. In this work, we revisit the longitudinal readout of the transmon qubit without using this significant approximation, instead accounting for the full cosine potential of the Josephson junction. This is particularly important in light of the advances made in the understanding of the phenomenology responsible for the non-QNDness of the dispersive readout~\cite{Dynamics_of_transmon_ionzation, dumas2024unified, MIST_1, MIST_2,Verney2019,Lescanne2019}. In those works, it was recognized that the loss of the QND character of the transmon is due to photon-number-specific resonances of the coupled qubit-resonator system~\cite{MIST_1}, and that the impact of these resonances can only be correctly captured by considering the cosine potential of the transmon. These resonances allow for population transfer to non-computational states, and are present due to the weakly anharmonic and multi-level nature of the transmon. Given that these two basic ingredients are also present in the longitudinal readout circuit of Ref.~\cite{Didier_longitudinal}, a natural question arises: could ionization limit the effectiveness of longitudinal readout?

In this work we demonstrate that this is not the case. Although these deleterious multi-photon transitions are also present in the longitudinal readout, we show that they are not as prevalent as in dispersive readout. Crucially, they can also be mitigated by  simple design choices. While previous work focused on the transmon in the two-level approximation~\cite{Didier_longitudinal}, our analysis and simulations keep the full cosine potential of the transmon~\cite{Koch_transmon}, which was shown to be crucially important in studying ionization~\cite{dumas2024unified}. As discussed in more detail below, simulations accounting for the full cosine potential suggest that longitudinal readout can achieve a readout fidelity of $99.99\%$ in less than \SI{50}{ns} with realistic circuit parameters. 

The rest of the manuscript is organized as follows: We first review the Hamiltonian of the circuit proposed in Ref.~\cite{Didier_longitudinal}  to realize longitudinal readout, and outline the general principles of the longitudinal transmon readout. We then use the tools discussed in Refs.~\cite{Dynamics_of_transmon_ionzation,Reminiscence_chaos,dumas2024unified} to analyze the ionization properties with branch analysis and through the Schrieffer–Wolff transformation. Subsequently, we discuss the effects of the selection rules of the longitudinal readout circuit, as well as the effect of disorder in the circuit. Taking into consideration the limitations posed by ionization, we then study the performance of longitudinal readout with realistic parameters. Furthermore, we compare the dispersive readout and longitudinal readout by treating both the transmon and the resonator classically, and making remarks on the chaotic behavior of their classical descriptions~\cite{Reminiscence_chaos}. Finally, we make concluding remarks. 

\section{Longitudinal readout in cQED} \label{sec:longitudinal_cqed}

\begin{figure}
    \centering
    \includegraphics[width=0.7\linewidth]{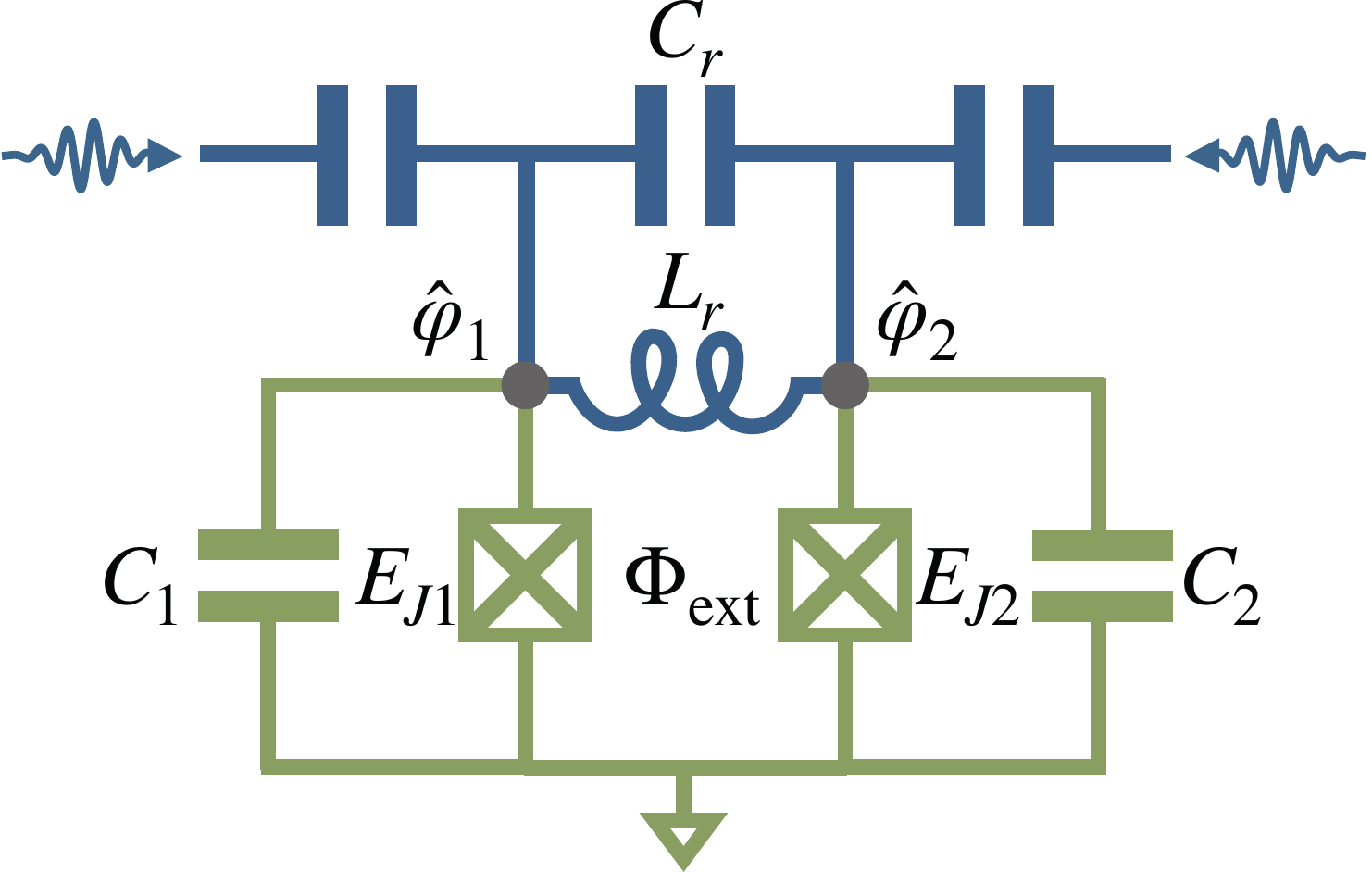}
    \caption{Circuit diagram for longitudinal readout of the transmon qubit. A flux tunable transmon (green) is inductively coupled to a resonator (blue). The symmetric $\hat{\varphi}_t = (\hat{\varphi}_1 + \hat{\varphi}_2)/2$ and antisymmetric $\hat{\varphi}_r = (\hat{\varphi}_1 - \hat{\varphi}_2)/2$ modes of the circuit correspond to the transmon mode and the resonator mode, respectively. The transmon and resonator modes can be driven by a symmetric (in-phase) and antisymmetric (out-of-phase) combinations of the two drives. In the text, we assume $C_1 = C_2$. See \cref{app_sec:circuit_quantization} for details.}
    \label{fig:longitudinal_circuit}
\end{figure}

The circuit realizing the longitudinal interaction is composed of a flux-tunable transmon inductively coupled to a readout resonator, see \cref{fig:longitudinal_circuit}~\cite{Didier_longitudinal}. Intuitively, the longitudinal interaction stems from the same mechanism that allows the control of the qubit frequency of a standard flux-tunable transmon \cite{Koch_transmon}. In the circuit of \cref{fig:longitudinal_circuit}, the differential mode $\hat{\varphi}_{r} = (\hat{\varphi}_1 - \hat{\varphi}_2)/2$ is the resonator degree of freedom which plays a similar role as the classical flux threading the loop of the tunable transmon. On the other hand, the common mode $\hat{\varphi}_t = (\hat{\varphi}_1 + \hat{\varphi}_2)/2$ is the qubit degree freedom. Expressed in terms of these degrees of freedom, the Hamiltonian of this circuit takes the form (see \cref{app_sec:circuit_quantization} for details) 
\begin{align}
    \hat{H} &= \omega_r \hat{a}^\dagger \hat{a} + 4 E_C (\hat{n}_t - n_g)^2 - E_J \cos{\hat{\varphi}_r}\cos{\hat{\varphi}_t} \nonumber \\ &- d E_J \sin{\hat{\varphi}_r}\sin\hat{\varphi}_t  -i \epsilon(t) \cos{(\omega_d t)} (\ha - \had),
    \label{eqn:Longitudinal_Hamiltonian_original}
\end{align}
where we have set $\hbar = 1$ and the external flux to zero, $\varphi_{\rm{ext}} = 0$. In this expression, $E_C$ is the transmon charging energy, $n_g$ the offset charge, $E_J = E_{J1} + E_{J2}$ the total Josephson energy with $d = (E_{J2} - E_{J1}) / 2$ the junction asymmetry. The resonator mode has frequency $\omega_r$, annihilation and creation operators $\hat{a}$ and $\hat{a}^\dagger$, and flux operator $\hat{\varphi}_r = \zpfr(\ha+\ha^\dagger)$, with $\zpfr = (2\pi/\Phi_0)\sqrt{Z_r/2}$ the resonator's zero-point phase fluctuations. Finally, $\omega_d$ is the drive frequency and $\epsilon(t)$ the time-dependent drive amplitude.

To better highlight the transmon-resonator interaction, we use a simple trigonometric identity to write \cref{eqn:Longitudinal_Hamiltonian_original} as
\begin{equation}\label{eqn:Longitudinal_hamiltonian}
    \begin{split}
        \hat{H} &= \omega_r \hat{a}^\dagger \hat{a} + \hat{H}_t -i \epsilon(t) \cos{(\omega_d t)} (\ha - \had)
        \\ &
        + 2 E_J \sin^2{(\hat{\varphi}_r / 2)} \cos{\hat{\varphi}_t}  
        -d E_J \sin{\hat{\varphi}_r} \sin{\hat{\varphi}_t}.
    \end{split} 
\end{equation}
The first line of this expression corresponds to the sum of uncoupled, driven cavity Hamiltonian and bare transmon Hamiltonian, $\hat{H}_t = 4 E_C(\hat{n}_t - n_g)^2 - E_J \cos \hat{\varphi}_t$. On the other hand, the second line corresponds to two types of transmon-resonator interactions. To gain an intuitive understanding of these interactions, it is useful to note that for the computational states of the transmon, $\cos\hvt$ is diagonal while $\sin\hvt$ is off diagonal. 

As a result, when projecting in the qubit subspace, the first term of the second line gives rise to a dispersive-like interaction, whereas the second term to a transverse interaction proportional to the junction asymmetry $d$, see \cref{app_sec:circuit_quantization}. The impact of finite junction asymmetry will be discussed in \cref{sec:asymmetry_&_disorder} where we will show its effects can be easily alleviated. We thus set $d=0$ for the moment.

Focusing on the first term of the second line of \cref{eqn:Longitudinal_hamiltonian}, we find that $\sin^2{(\hat{\varphi}_r / 2)} $ has off-diagonal matrix elements which can result in transitions between Fock states. Crucially, it also has diagonal matrix elements in the Fock basis which gives rise to a dispersive interaction. Expressing the Hamiltonian in terms of its diagonal and off-diagonal components---something which can be done exactly---and projecting the transmon onto its two lowest energy levels yields (see \cref{app_sec:SW} for details)
\begin{equation}\label{eqn:two_level_approx_H}
    \begin{split}
        \hat{H}
        &=
        [
        \omega_r^\prime 
        +\chi_z(\hat{a}^\dagger \hat{a}) \hat{\sigma}_z
        ]\hat{a}^\dagger \hat{a} 
        +
        \frac{\omega_q}{2} \hat{\sigma}_z
        +
        \hat{V}(\ha^2, \ha^{\dagger 2})
        \hat{\sigma}_z
        \\ 
        &
        -i \epsilon(t) \cos{(\omega_d t)} (\ha - \had),
    \end{split}
\end{equation}
where $\omega_r'$ is the dressed resonator frequency and $\omega_q \approx \omega_p - E_C$ is the qubit frequency with $\omega_p  = \sqrt{8 E_J E_C}$ the transmon's plasma frequency. To second order in $\zpfr$, $\chi_z(\had \ha) \approx \omega_p \zpfr^2/4$ 
and the first term of $\hat{H}$ takes the form of the usual dispersive coupling. On the other hand, $\hat{V}(\ha^2, \ha^{\dagger 2})$ 
is not diagonal in the Fock basis but can only cause transitions between Fock states of the same parity. Crucially, in contrast to the dispersive readout, here the dispersive interaction is not derived perturbatively from a parent transverse interaction. The full non-perturbative forms of $\chi_z(\had \ha)$ and $\hat{V}(\ha^2, \ha^{\dagger 2})$ are indeed simply related to the matrix elements of the displacement operator in the Fock basis~\cite{Cahill_Glauber}, and their exact expressions can be found in \cref{app_sec:SW}. Because this is not a perturbative result, the dispersive interaction strength does not scale with the inverse of the qubit-resonator detuning as it does when it originates from a transverse interaction. 

In Ref.~\cite{Didier_longitudinal}, the longitudinal interaction $\hH_z$ is obtained using the circuit of \cref{fig:longitudinal_circuit} by modulating the external flux threading the qubit loop at the resonator frequency. Here, we use an alternative approach which, as discussed above, instead relies on driving the resonator~\cite{Mathieu_thesis}. This is made clear by moving to a displaced frame $\ha \to \ha + \alpha(t)$ to eliminate the drive from the Hamiltonian, where $\alpha(t)$ is determined by $\epsilon(t)$ via the linear response properties of the resonator.  Momentarily ignoring the off-diagonal contribution $\hat{V}(\ha^2, \ha^{\dagger 2})$ as well as the nonlinearity of the $\chi_z$ shift, in this new frame the transformed Hamiltonian $\hat{H}'$ takes the form~\cite{Blais2007}
\begin{equation}
    \hat{H}^\prime
    =
    [\omega_r + \chi_z(0) \hat{\sigma}_z] \hat{a}^\dagger \hat{a}
    +
    \frac{\omega_q}{2}\hat{\sigma}_z 
    +
    \chi_z(0) \alpha(t) (\ha^\dagger+\ha)\hat{\sigma}_z.
    \label{eqn:activate_longitudinal_interaction}
\end{equation}
The last term of $\hat{H}'$ has the desired longitudinal interaction of amplitude $g_z(t) = \chi_z(0) \alpha(t)$ proportional to the magnitude of the resonator displacement. At large displacement amplitudes, the longitudinal interaction overwhelms the dispersive-like interaction of amplitude $\chi_z(0)$ also present in $\hat{H}'$. To turn on the longitudinal interaction, below we use a measurement drive amplitude $\epsilon(t)$ which rapidly fills the cavity with photons, see \cref{app_sec:pulse_shape} for details on our choice of pulse shape. 

This approach to obtain a longitudinal interaction using a microwave drive rather than a flux modulation is reminiscent of the experiments reported in Refs.~\cite{Touzard_synthetic_long, Ikonen_synthetic_long}. There, a longitudinal interaction is produced by driving a transversally coupled qubit with a frequency matching that of the resonator, reminiscent of the mechanism leading to the cross-resonance gate~\cite{Chow2011}. A crucial difference is that because it is based on a transverse qubit-resonator coupling, this approach is susceptible to the same non-QNDness as the dispersive readout. Moreover, for the same reason, the dispersive shift $\chi_x$ and thus the effective longitudinal coupling $g_z \propto \chi_x$ scales inversely with the qubit-resonator detuning.

In addition to those mentioned in \cref{sec:intro}, an advantage of the longitudinal readout is the lack of Jaynes-Cummings critical photon number $n_\mathrm{crit}^\mathrm{jc} = (\Delta/2g)^2$ present in the standard dispersive readout, where $\Delta = \omega_q - \omega_r$~\cite{RMP}. This quantity is associated with the breakdown of the dispersive approximation. As the Hamiltonian in \cref{eqn:activate_longitudinal_interaction} is not derived using the dispersive approximation, there are no directly analogous quantities here. Indeed, although \cref{eqn:activate_longitudinal_interaction} was obtained from \cref{eqn:Longitudinal_hamiltonian} under two approximations --- projecting the Hamiltonian onto the computational subspace and ignoring $\hat{V}(\ha^2, \ha^{\dagger 2})$ --- these seem well justified. Within perturbation theory, any transitions we ignored  are highly suppressed: the detuning between any such transitions is large and any $2n$-photon transition amplitude will be multiplied by a factor $(\zpfr/2)^{2n} \ll 1$. It thus seems as though \cref{eqn:activate_longitudinal_interaction} remains valid even in the presence of strong drives. We explore this question in more details in the next section.

\section{Ionization in the longitudinal readout} 

Transmon ionization, or measurement-induced state transitions, in the dispersive readout is a consequence of photon-number-specific accidental degeneracies in the coupled qubit-resonator system~\cite{Dynamics_of_transmon_ionzation, dumas2024unified, MIST_1, MIST_2, Reminiscence_chaos, Xiao2023}. Despite the QND character of the simplified model of \cref{eqn:two_level_approx_H}, such degeneracies are possible in the full Hamiltonian \cref{eqn:Longitudinal_hamiltonian} of the circuit of \cref{fig:longitudinal_circuit}. In this section, we use perturbative arguments as well as the tools developed in Refs.~\cite{Dynamics_of_transmon_ionzation, dumas2024unified} to investigate transmon ionization in the longitudinal readout with the circuit of \cref{fig:longitudinal_circuit}. More specifically, we use branch analysis to identify ionizing critical photon numbers $n_{\rm crit}$, and demonstrate that qubit-resonator resonances are not as detrimental when compared to dispersive readout. Moreover, we show that the critical photon numbers can be made larger simply by increasing the transmon-resonator detuning. Crucially, because of the non-perturbative nature of the cross-Kerr interaction in this circuit, this increase in detuning can be done without affecting the dispersive shift and thus the longitudinal coupling $g_z$. We then show that nonidealities in the longitudinal circuits, such as gate charge and junction asymmetry, do not significantly degrade its robustness to ionization.

\subsection{Physical origin of ionization and simple mitigation strategies}

To gain an intuitive understanding of the physical mechanism leading to ionization in the longitudinal readout, we begin by approximately diagonalizing the full undriven circuit Hamiltonian \cref{eqn:Longitudinal_hamiltonian}. We first express the Hamiltonian as $\hH = \hH_0 + \hat{V}$, separating it into its diagonal component $\hH_0$ and off-diagonal component $\hat{V}$. Without approximations, the diagonal part can be written as 
\begin{equation}\label{eq:H_0}
    \begin{split}
        \hH_0
        &=
        \sum_{j_t}
        \left[
        \omega_r + \chi_{z, j_t}(\ha^\dagger \ha)
        |j_t\rangle \langle j_t|
        \right]
        \hat{a}^\dagger \hat{a}
        \\ 
        &+
        \sum_{j_t}
        \left(\omega_{j_t} + \Lambda_{j_t} \right)
        |j_t \rangle \langle j_t|,
    \end{split}
\end{equation}
with $\omega_{j_t}$ the bare transmon energies and $\ket{j_t}$ the bare transmon states. The exact expressions for the Lamb shifts $\Lambda_{j_t}$ and the nonlinear and non-perturbative dispersive shifts $\chi_{z,j_t}(\ha^\dagger \ha)$ can be found in \cref{app_sec:SW}.

The perturbation $\hat{V}$ is then, by definition, the off-diagonal part of the interaction $2 E_J \sin^2(\hat{\varphi}_r/2) \cos \hat{\varphi}_t$. Although the corresponding matrix elements can be computed exactly, for the purposes of simplifying the discussion we make two approximations. We stress that these approximation are not made in the numerical results presented below. First, we treat the transmon as a Kerr nonlinear oscillator with $\hat{n}_t = -in_{{\rm zpf}, t}(\hat{b}-\hat{b}^\dagger)$, $\hat{\varphi}_t = \varphi_{{\rm {zpf}}, t}(\hat{b} + \hat{b}^\dagger)$, where $n_{{\rm zpf}, t}$ and $\zpft$ are the zero-point fluctuations of the charge and phase, respectively~\cite{RMP}. 
Second, we only keep terms to the lowest non-trivial order in $\zpfr$ and $\zpft$, leading to \cite{Mathieu_thesis}
\begin{equation}\label{eq:V_perturbative}
    \begin{split}
        \hat{V}
        \approx
        \frac{E_J \zpfr^2}{2}(\ha+\ha^\dagger)^2
        \left(
        1-\frac{\zpft^2}{2}(\hb+\hb^\dagger)^2
        \right)
        \Bigg|_{\rm off-diag},
    \end{split}
\end{equation}
where diagonal terms such as $\ha^\dagger \ha \hb^\dagger \hb $ have already been taken into account in $\hH_0$. To this order in perturbation theory, $\hat{V}$ can create or destroy pairs of excitation via terms such as  $\ha^\dagger \ha(\hb^{2} + \hb^{\dagger 2})$. These processes are, however, highly off resonant, and can safely be eliminated using the standard rotating-wave approximation (RWA). The interaction $\hat{V}$ also allows the transmon and resonator to exchange pairs of excitation via terms such as $\ha^{\dagger 2} \hb^2 + \hb^{\dagger 2} \ha^2$. Because the qubit and resonator are taken to be detuned from each other, these processes are off resonant in the absence of a drive. The nonlinear dispersive shift $\chi_{z, j_t}(\ha^\dagger \ha) |j_t \rangle \langle j_t| \ha^\dagger \ha$ present in \cref{eq:H_0} can make such transitions resonant in the presence of resonator photons, and we thus expect such transitions to be the dominant factor leading to ionization. Crucially, however, these two-excitation exchange terms can be mitigated by simply increasing the qubit-cavity detuning. As a result, more photons are needed to make these unwanted processes resonant and the onset of ionization is pushed to higher photon numbers. As explained above, this increase in the detuning can be made without sacrificing the dispersive interaction. This is in stark contrast to the dispersive readout, where the amplitude of the dispersive shift $\chi_x$ is inversely proportional to the detuning~\cite{RMP}.

Another subtle advantage of this coupling scheme is the presence of a discrete symmetry. Indeed, formally the full undriven Hamiltonian has a $\mathbb{Z}_2 \times \mathbb{Z}_2$ symmetry, as it is invariant under either parity transformations $\h{\varphi}_r \to -\h{\varphi}_r$ and $\h{\varphi}_t \to -\h{\varphi}_t$. While to the lowest order in perturbation theory $\hat{V}$ only involves pairs of excitation on either the transmon or resonator, the presence of this symmetry ensures that this conclusion is valid to all orders. This is unlike a standard transverse coupling, which only preserves the total excitation parity. This additional symmetry results in fewer pathways for the qubit and resonator to hybridize \cite{Lu_parity_protected, Mathieu_thesis}, and we thus expect our circuit to be more robust to ionization than the standard dispersive readout. Although this symmetry is broken for a finite junction asymmetry $d \neq 0$, such a term would be made negligible by increasing the qubit-resonator detuning for the same reason as mentioned above, see \cref{sec:asymmetry_&_disorder}.

From this perturbative analysis, we can thus expect that i)~the enhanced parity symmetry will prevent unwanted mixing between the transmon and resonator states, ii)~the dressed states will closely resemble the bare counterparts, at least until ionization sets in, and iii)~ionization can be pushed to higher photon number by increasing the resonator-transmon detuning.

\begin{figure*}[ht]
    \centering
    \includegraphics[width=\linewidth]{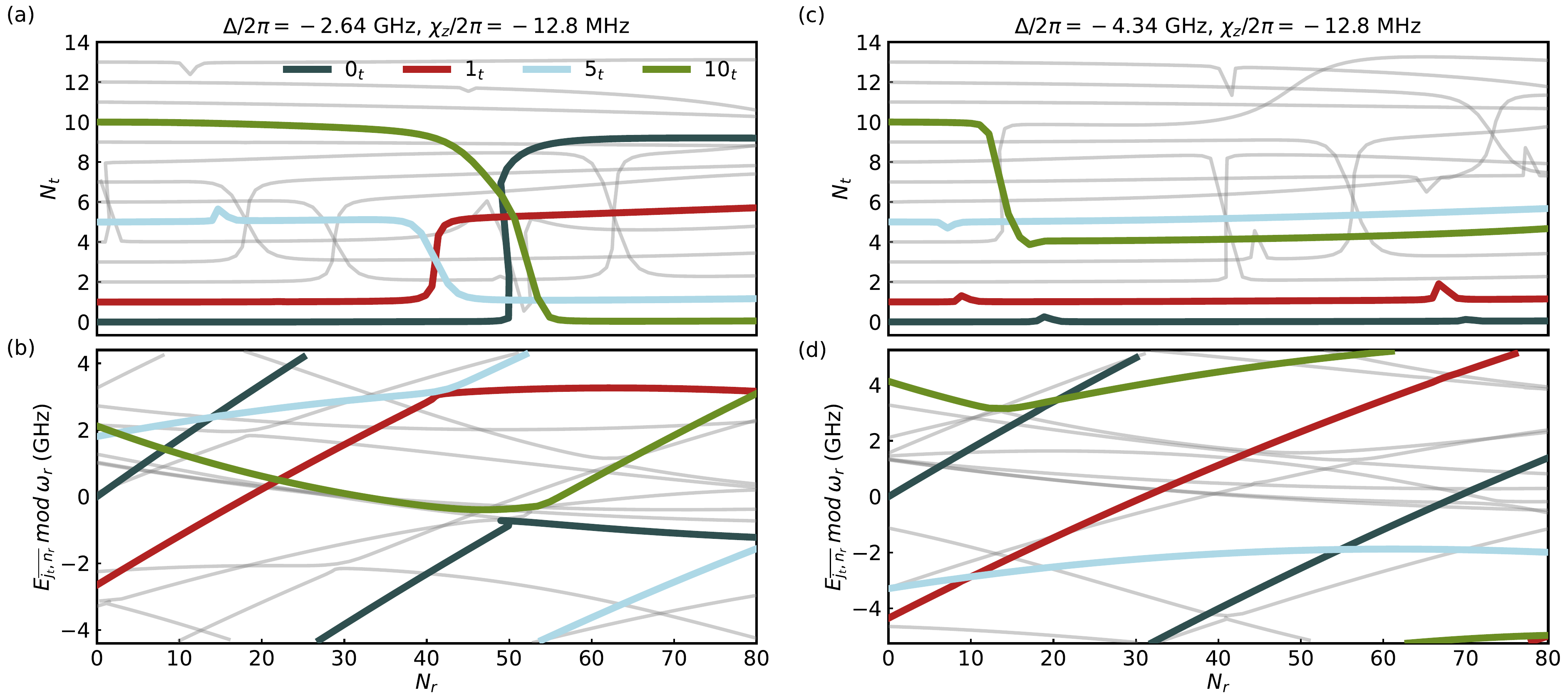}
    \caption{(a) Transmon population as a function of the average resonator population. The ground and excited state branches, as well as the branches they swap with are highlighted. The transmon parameters are $E_C / 2 \pi = 0.215$ GHz, $E_J / E_C = 110$, with junction asymmetry $d = 0$ and gate charge $n_g = 0$. The resonator frequency is $\omega_r / 2 \pi = 8.8$ GHz. We keep 16 states in the transmon, and 100 states in the resonator. The qubit-resonator detuning is defined as $\Delta = \omega_q - \omega_r$. (b)  The energy modulo $\omega_r$ spectrum as a function of the average resonator population. Parameters are the same as in (a). The wrapping of the energies is due to the spectrum being a modulo of the resonator frequency, indicating that when two lines cross, the two states are energetically separated by an integer number of resonator photons. Avoided crossings align with branch swappings occuring in (a). (c) Transmon population as a function of average resonator population with all parameters the same as in (a) but with $\omega_r / 2 \pi = 10.5$ GHz. The branch swappings seen in (a) are now removed due to the higher detuning between the resonator and the transmon. (d) Energy modulo $\omega_r$ spectrum for the parameters of panel (c). Importantly, the dispersive shift $\chi_z$ is the same in (a) and (c) as it is independent of detuning.}
    \label{fig:branch_analysis_1}
\end{figure*}

\subsection{Branch analysis} \label{sec:branch_analysis}

Having obtained an intuitive understanding of ionization in the longitudinal readout, we now turn to exact numerical calculations to confirm our predictions. To do so, we diagonalize the full undriven Hamiltonian \cref{eqn:Longitudinal_hamiltonian}, and use the branch analysis to label the eigenstates of the composite undriven qubit-resonator system to obtain photon-number-dependent transmon branches~\cite{Dynamics_of_transmon_ionzation, dumas2024unified, Boissonneault_qubit_readout}. Each of these branches contains information about the transmon’s energy levels as they vary with the photon number in the resonator. This information can subsequently be used to understand the sudden population transfer, or ionization, of the transmon in the presence of a drive. Before doing this, we first outline the basic principle and construction of the branch analysis, see Ref.~\cite{Dynamics_of_transmon_ionzation,dumas2024unified} for further details. The first step in the construction of the branches is to numerically diagonalize the full undriven Hamiltonian \cref{eqn:Longitudinal_hamiltonian} in the eigenbasis $|j_t, n_r\rangle$ of the uncoupled transmon-resonator system resulting in the set of eigenvectors $\{ | \lambda \rangle \}$. For each $j_t$ we then find the state $| \lambda \rangle$ which maximizes the overlap $|\langle j_t, 0_r | \lambda \rangle |^2$ and label it $\ket{\overline{j_t, 0_r}}$. We then use these states $|\overline{j_t, 0_r}\rangle$ to recursively assign a label to all remaining states: $\ket{\overline{j_t, n_r + 1}}$ is defined as the unassigned eigenstate $|\lambda \rangle$ which maximizes the overlap $|\langle \lambda | \hat{a}^\dagger | \overline{j_t, n_r} \rangle|^2$. Repeating this for each transmon index $j_t$ results in a set of dressed states $B_{j_t} = \{ \ket{\overline{j_t, n_r}} \}$ which we refer to as branches. 

In \cref{fig:branch_analysis_1}(a) we show a parametric plot of the transmon population $N_t(\overline{j_t, n_r}) = \sum_{k_t, m_r} k_t \vert \langle k_t, m_r \vert \overline{j_t, n_r} \rangle \vert^2$ of each branch as a function of the average photon number $N_r(\overline{j_t, n_r}) = \langle \overline{j_t , n_r} | \ha^\dagger \ha |\overline{j_t, n_r} \rangle$. We see that at certain photon numbers, different branches ``exchange" their transmon population, a process referred to as branch swapping~\cite{dumas2024unified}. Branch swapping at a given $N_r$ result from the transmon-resonator coupling in \cref{eqn:Longitudinal_hamiltonian} and have a simple dynamical consequence: population transfer to the corresponding transmon state can occur when this number of photons are placed in the resonator. These swapping thus lead to transition out of the computational subspace and limit the QND nature of readout~\cite{dumas2024unified,Dynamics_of_transmon_ionzation}. 

These branch swappings arise due to resonant multi-photon transitions between two dressed transmon states~\cite{dumas2024unified,Dynamics_of_transmon_ionzation}. To illustrate this we show in \cref{fig:branch_analysis_1}(b) the eigenenergies of the first $15$ branches modulo the resonator frequency $\omega_r$. Energetic collisions between two states in this figure means that they are resonant via some $m$-photon transition \cite{MIST_1}. For each branch swapping in \cref{fig:branch_analysis_1}(a) there is an accompanying avoided crossing in the modular energy spectrum of \cref{fig:branch_analysis_1}(b). In some cases, branches cross without an avoided crossing and therefore without branch swapping. This is due to the magnitude of the transmon-resonator coupling matrix element which scales with $\sqrt{n_r}$. At low photon number, these matrix elements can be too small to connect widely separated transmon states. For example, the crossing of $B_{0_t}$ and $B_{10_t}$ at $N_r \sim 8$ photons in \cref{fig:branch_analysis_1}(b) does not lead to a branch swapping in \cref{fig:branch_analysis_1}(a). At a larger photon number, $N_r \sim 50$, the same two branches collide again but show an anticrossing and a corresponding branch swap. 

The branch analysis --- which is by construction non-perturbative --- confirms the validity of the intuition built using perturbation theory. For instance, unlike the standard dispersive coupling, we note that there is never any swapping between branches of different parity, a consequence of the $\mathbb{Z}_2 \times \mathbb{Z}_2 $ symmetry. Consequently, there are several crossings (as opposed to avoided crossings) between branches of different parity in the modular energy spectrum. This is in stark contrast to dispersive readout, which emerges from a charge-charge coupling and therefore connects the two parity sectors of the transmon. In addition, before ionization, we see that the character of the qubit's computational branches are largely unaffected by the presence of photons. Indeed, as we see in \cref{fig:branch_analysis_1}(a), the transmon populations of either branch $N_t(\overline{0_t, n_r}) \approx 0$ and $N_t(\overline{1_t, n_r}) \approx 1$ are approximately independent of photon number, indicating that these dressed states closely resemble their bare counterparts. Once again, this is to be contrasted with the standard dispersive readout, in which the transmon populations increase linearly with photon number $N_t(\overline{0_t, n_r}) \propto n_r$ and $N_t(\overline{1_t, n_r}) \propto n_r$, a consequence of the weak transmon-resonator hybridization.

In \cref{fig:branch_analysis_1}(a,c) we compare the results of the branch analysis for two different qubit-resonator detunings of (a) $\Delta/2\pi = -2.64$ GHz and (c) -4.34 GHz at a constant dispersive shift $\chi_z / 2 \pi = -12.8$ MHz. As expected from the above discussions, the critical photon number can be made larger by increasing the qubit resonator detuning, without affecting the value of $\chi_z$ and thus of the longitudinal coupling $g_z$. As discussed in more detail in \cref{sec:readout_performance}, we stress that for a wide range of parameters, the critical photon numbers associated to ionization are well above the number of photons required to do fast readout. Increasing the qubit-resonator detuning is thus a valid strategy to increase the ionization threshold for a wide range of transmon parameters. 

\begin{figure}
    \centering
    \includegraphics[width=\linewidth]{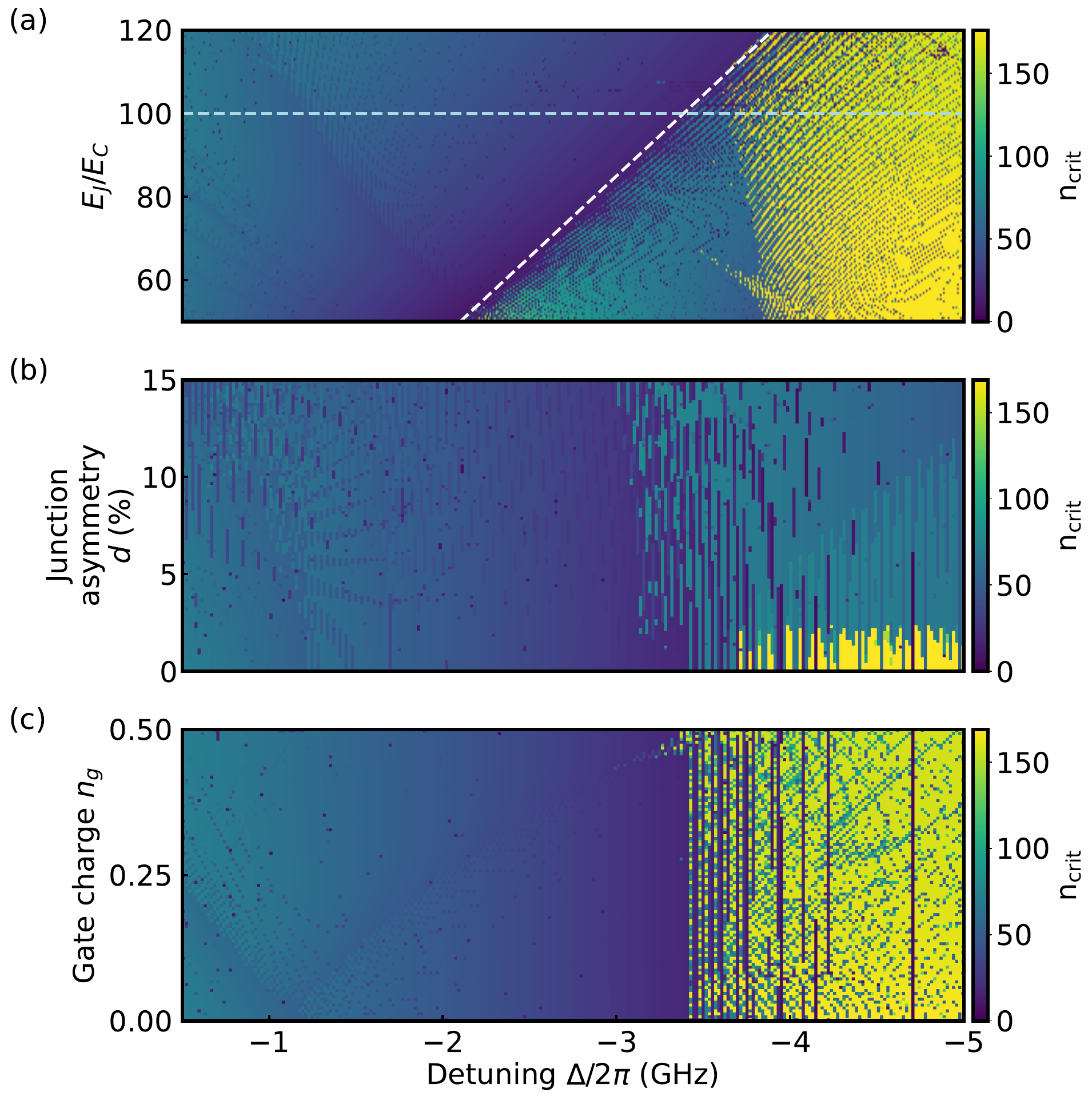}
    \caption{(a) Critical photon numbers as a function of $E_J/E_C$ and transmon-resonator detunings $\Delta = \omega_q - \omega_r$. The diagonal white line indicates a strong resonance between the first excited and the fifth excited state of the transmon that spans across all values of $E_J/E_C$. The blue line indicates $E_J/E_C = 100$, corresponding to the fixed $E_J / E_C$ ratio used in panels (b) and (c). (b) Ionization critical photon numbers for various detunings $\Delta / 2\pi$ and Josephson junction asymmetries $d$. (c) Ionization critical photon numbers as a function of detuning $\Delta / 2\pi$ and gate charge $n_g$. For finite asymmetry and non-zero gate charge, the $\mathbb{Z}_2 \times \mathbb{Z}_2$ symmetry is broken. Despite this, we observe that the critical photon numbers, especially at high detunings remain large. $E_J/E_C = 100$ in panels (b) and (c)}
    \label{fig:ncrits_v_detuning}
\end{figure}

To illustrate that this conclusion applies beyond the parameters of \cref{fig:branch_analysis_1}, in \cref{fig:ncrits_v_detuning}(a) we plot the critical photon numbers as a function of the $E_J/E_C$ ratio and qubit-resonator detuning $\Delta = \omega_q - \omega_r$. Here, the critical photon numbers are defined as the minimum of the critical photon numbers for the branches $B_{0_t}$ and $B_{1_t}$, i.e.~$n_{\rm{crit}} = \textrm{min}\{n_{\rm{crit, 0_t}}, n_{\rm{crit, 1_t}}\}$. We observe a diagonal region of low critical photon numbers beginning at $\Delta / 2\pi \approx -2$ GHz for $E_J / E_C = 50$, which persists across all $E_J/E_C$ values, corresponding to a strong two photon resonance between the first and fifth excited state of the transmon, see white dashed line. Indeed, this region of low critical photon number occurs due to the constant dispersive shift of the interaction, and is dependent on how far the transmon needs to be ac-Stark shifted in order to cause an accidental degeneracy with a higher state. By increasing the detuning, degeneracies with the computational states can be avoided as the energy difference between the different transmon states modulo the resonator frequency increases, thus pushing the multi-photon transition further away. Indeed, the critical photon numbers increase dramatically once the detuning is increased a little past the low critical photon number region. The above phenomenon also persists when the junction asymmetry and gate charge are varied, see \cref{fig:ncrits_v_detuning}(b) and (c). Once again, the low critical photon number region is due to the same frequency collisions between the first and fifth excited states of the transmon at low photon numbers.

Finally, \cref{fig:ncrits_v_detuning}(a) illustrates that for any value of $E_J / E_C$, there exists regions where the critical photon numbers are well above the number of photons required for a fast readout. As is expected from \cref{fig:branch_analysis_1}(a), if we further increase the detuning between the transmon and the resonator the critical photon numbers increase for all values of $E_J/E_C$. 

\subsection{Asymmetries in the Josephson junctions / Effect of gate charge} \label{sec:asymmetry_&_disorder}

In the above discussion, we have taken the two Josephson junctions to be identical. Realistic devices will have some amount of junction asymmetry ($d \neq 0$) and, in that case, the last term of \cref{eqn:Longitudinal_hamiltonian} in the two-level projection results in a transverse interaction term between the transmon and the resonator of the form $g_x (\hat{a}^\dagger + \hat{a})\hat{\sigma}_x $. This interaction lifts the selection rules suppressing ionization, allowing for more transition to occur. While junction asymmetry should therefore be minimized in the circuit of \cref{fig:longitudinal_circuit}, the adverse effects of this interaction can be mitigated by increasing the detuning between the transmon and the resonator. Following the discussion from the previous section, this is a second reason to work at large transmon-resonator detuning.

\cref{fig:ncrits_v_detuning}(b) shows the critical photon numbers for varying junction asymmetry up to $15$\%. We observe for most detunings $\Delta$, the critical photon numbers do not vary substantially within a few percent of asymmetry. Especially at the higher detuning where the critical photon numbers are high, we see a sudden drop off in the critical photon numbers. However, it is important to point out that the critical photon numbers are still high and well above the number required for a fast readout. The sensitivity to asymmetry is most apparent where the critical photon numbers are low to begin with, suggesting that the branch swappings that occur in those regions move around due to the energy shift caused by the junction asymmetry. How tolerant the parameter regime is to asymmetry will vary for the choice of $E_J/E_C$ and the resonator frequency. It is then paramount to not only choose the parameters that grant a high critical photon number in ideal conditions, but to also choose parameters that are tolerant against some imperfection. 

In addition to being robust to junction disorder $d$, the critical photon number of the longitudinal readout only weakly depends on the transmon's gate charge $n_g$. This is to be contrasted with the dispersive readout where changes in the gate charge result in large variations of the critical photon number \cite{dumas2024unified,Reminiscence_chaos, MIST_2}. This robustness is illustrated in \cref{fig:ncrits_v_detuning}(c) which shows for $E_J / E_C = 100$ the critical photon number as a function of qubit-resonator detuning and gate charge in the interval $n_g \in [0,0.5]$. 
The overall features that are observed in panel (b) are similar to what we observe in panel (a) where the critical photon numbers remain high at large detunings. In short, both non-idealities $d$ and $n_g$ have a small effect on the critical photon number.  

\section{Fast and high-fidelity longitudinal readout} \label{sec:readout_performance}

\begin{figure}
    \centering
    \includegraphics[width=\linewidth]{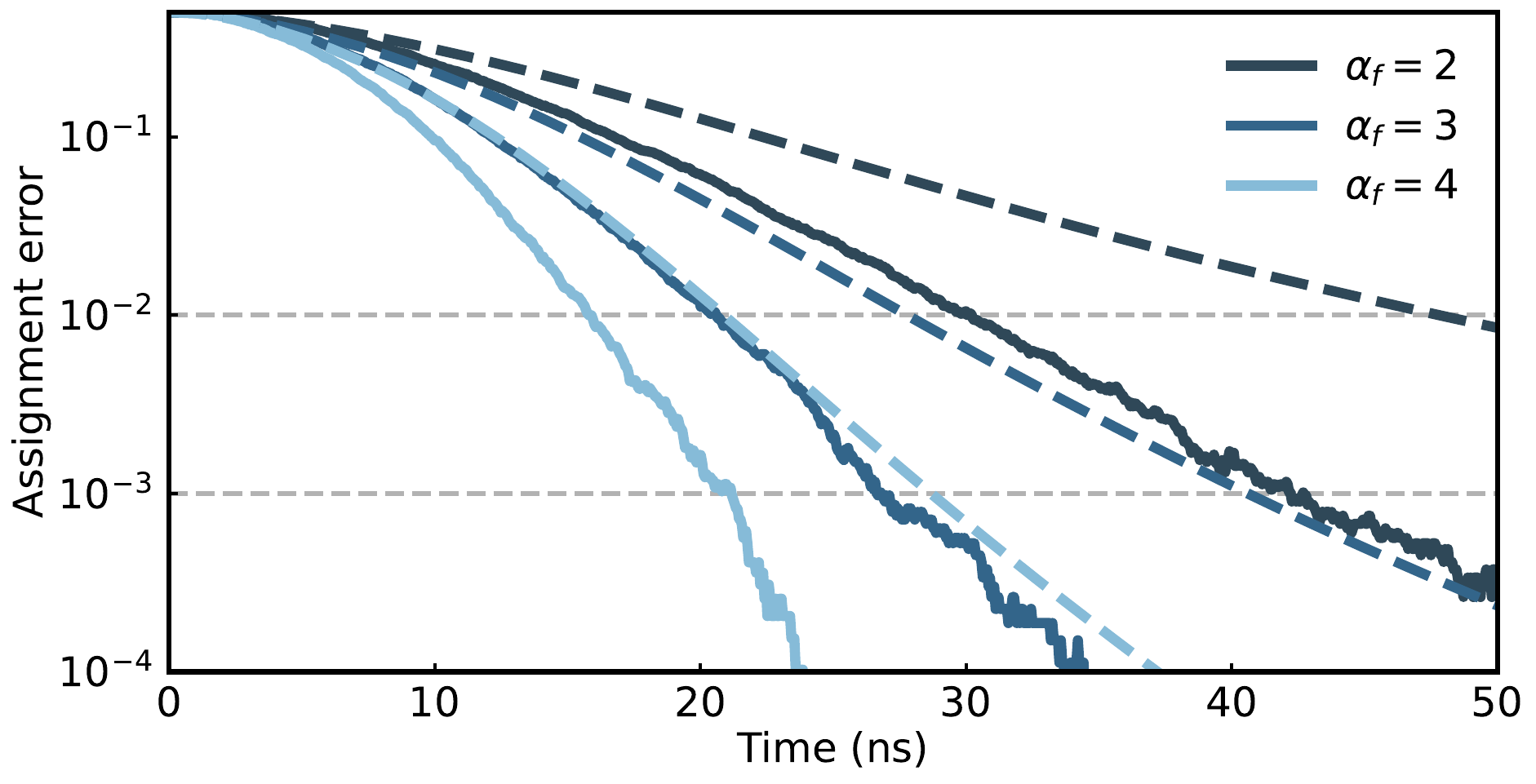}
    \caption{Readout assignment error as a function of integration time $\tau$ for drive strengths corresponding to $\alpha_f = 2, 3$, and $4$. The parameters are $E_J / E_C = 50$, $d = 0$, $n_g = 0$, $\omega_r / 2 \pi = 9.3$ GHz, $\kappa / 2 \pi = 17$ MHz and qubit relaxation is ignored. The dispersive shift is $\chi_z / 2 \pi \approx -8.66$ MHz and the qubit frequency $\omega_q/2\pi \approx 4.07$ GHz, corresponding to $\Delta / 2 \pi = -5.3$ GHz. Solid lines are results obtained from heterodyne readout simulations with the measurement efficiency set to $\eta = 1$. Dashed lines are computed using the analytical expression for the signal-to-noise ratio with efficiency set to $\eta = 0.5$.} See \cref{app_sec:pulse_shape} and \cref{app_sec:stochastic_readout_sims} for further details on the readout simulations.
    \label{fig:RO_sim}
\end{figure}

As discussed in \cref{sec:intro}, under the longitudinal interaction $g_z(\hat{a}^\dagger + \hat{a})\hat{\sigma}_z $, the qubit-state dependent displacement of the resonator moves at the optimal $180$ degrees out of phase from one another. Therefore, especially at shorter integration times, the signal-to-noise ratio increases more rapidly than for the dispersive readout as the pointer states separate in the optimal manner~\cite{Didier_longitudinal}. To quickly activate the longitudinal interaction and obtain a fast and high-fidelity readout, we begin by quickly displacing the resonator field to a target coherent state $\ket{\alpha_f}$ leading to a longitudinal interaction of amplitude $g_z = \alpha_f\chi_z$, see \cref{app_sec:pulse_shape} for the pulse shape $\varepsilon(t)$. As an illustration, \cref{fig:RO_sim} shows the assignment error for different $\alpha_f$ with $E_J / E_C = 50$ and $\omega_r / 2 \pi = 9.3$ GHz, obtained from stochastic Schr\"odinger equation simulations of the readout dynamics with heterodyne detection. Note that here we use $E_J/E_C = 50$ as opposed to $E_J/E_C = 100$ as in \cref{fig:branch_analysis_1}, as this gives a larger $\zpft$ and consequently a larger dispersive shift $\chi_z$. However, a large $E_J/E_C$ ratio does not limit the readout performance. 

Even for a moderate driving amplitude corresponding to $|\alpha_f|^2 \sim 16$ photons (light blue), the assignment error reaches $10^{-4}$ in less than $30$ ns for a measurement efficiency of $\eta = 1$, and in less than $50$ ns for a more realistic measurement efficiency of $\eta = 0.5$ (dashed light blue). Those results are obtained for photon numbers that are well below the critical photon number, here $\sim 170$ photons for the chosen parameters. See \cref{app_sec:stochastic_readout_sims} for further details on the stochastic simulations. 

Ideally, we want the fidelity and QNDness of the readout to be maximized. The two processes affecting these metrics are ionization and the resonator's self-Kerr. At high photon numbers, the resonator's self-Kerr --- which just like the dispersive shift $\chi_z$ is non-perturbative and not controlled by the qubit-resonator detuning ---can lead to bananization of the resonator field~\cite{Sivak2019,Boutin2017}. This can reduce the signal-to-noise ratio and thereby reduce the readout fidelity. Furthermore, a large self-Kerr can limit the maximum number of photons that can be placed into the resonator. Despite these possible limitations to the readout, \cref{fig:RO_sim} which contains these imperfections shows that fast and high-fidelity longitudinal readout is possible.

\section{Emergence of chaos in the longitudinal readout}

\begin{figure*}
    \centering
    \includegraphics[width=\linewidth]{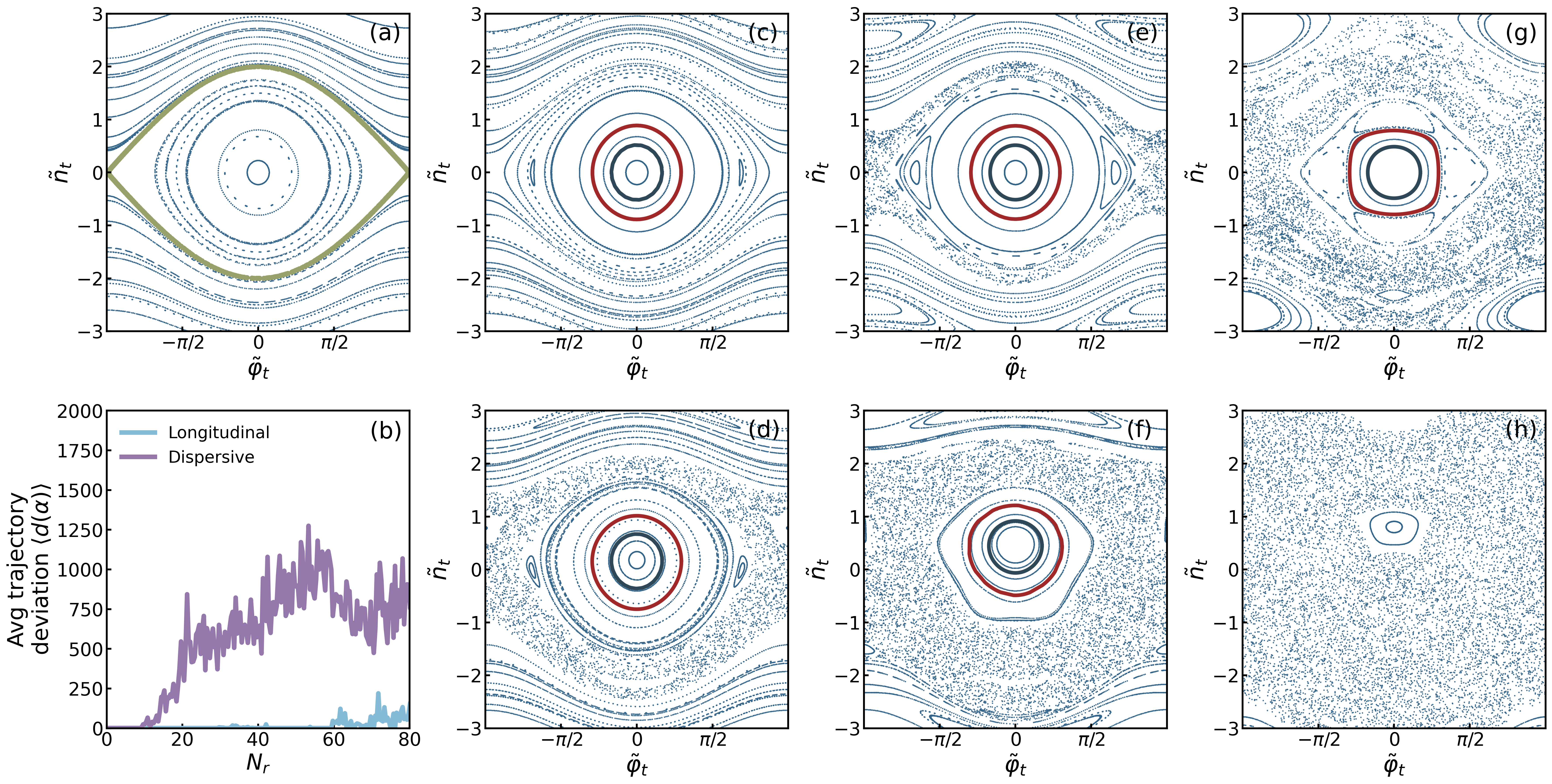}
    \caption{(a) Poincare section for a transmon with no drive. The green line indicates the separatrix. (b) The average trajectory deviation with increased drive amplitude. The average trajectory deviation is calculated by averaging over many trajectory deviations calculated from \cref{eqn:chaos_trajectory_deviation}. (c), (e), and (g) show the Poincare section for a transmon coupled via the longitudinal interaction with drive strengths corresponding to $1$, $9$, and $49$ photons, respectively. The blue and red orbits indicate the Bohr-Sommerfeld orbits for the ground and first excited state of the transmon, respectively. (d), (f), and (h) show Poincare sections for a transmon coupled dispersively to a resonator again with $1$, $9$, and $49$ photons, respectively. The blue and red orbits indicate the Bohr-Sommerfeld orbits for the ground and first excited state of the transmon, respectively. In panel (h), the regular region is too small to support a Bohr-Sommerfeld orbits. For both longitudinal and dispersive readout cases,  $E_J / E_C$ of the transmon was set to 110. For longitudinal readout the asymmetry was set to $d = 0$.}
    \label{fig:chaos}
\end{figure*}

In the previous sections, we have shown that the critical photon number of the longitudinal readout can be made larger than in a comparable dispersive readout, and have argued that this robustness is linked to the symmetry of the qubit-resonator coupling in the circuit of \cref{fig:longitudinal_circuit}. In this section, we provide another qualitative argument for the robustness of the longitudinal readout and its advantage over the dispersive readout based on the emergence of chaos in the classical model of those two readouts~\cite{Reminiscence_chaos}. To do so, we follow the approach of Ref.~\cite{dumas2024unified} where it was shown that the presence of classical chaos in the driven classical transmon can be linked to transmon ionization in the dispersive readout.

First, in the case of the usual dispersive readout, the classical model of the transmon is obtained from the quantum transmon-resonator Hamiltonian by neglecting the quantum fluctuations of the resonator and effectively treating the driven resonator as a classical drive on the transmon. The resulting Hamiltonian is that of a classical pendulum driven by a monochromatic drive on the pendulum's conjugate momentum \cite{Graham1991Level,Reminiscence_chaos,dumas2024unified} with Hamiltonian
\begin{equation}
    \tilde{H} = \frac{1}{2}\tilde n_t^2-\cos\tilde{\varphi}_t + \varepsilon_t \cos(\tilde\omega_d\tilde t)\tilde n_t.
\end{equation}
Here, $\Tilde{H} = H / E_J$, $\tilde \varphi_t = \varphi_t$, $\tilde n_t = z n_t$, and $z = \sqrt{8 E_C/E_J}$ is the transmon's impedance. The phase space coordinates satisfy the modified Poisson bracket $\{\tilde\varphi_t,\tilde n_t\}=z$. Moreover, time is rescaled to $\Tilde{t} = \omega_p t$ where $\omega_p = \sqrt{8 E_J E_C}$ is the plasma frequency. As a result,  the renormalized drive frequency is $\Tilde{\omega}_d = \omega_d / \omega_p$. Following the same approach for the longitudinal readout results in the replacement $\hat{\varphi}_r \to 2\varepsilon_p \cos{(\omega_d t)}$ in \cref{eqn:Longitudinal_hamiltonian} with $\varepsilon_p = \zpfr \alpha_f$. In this expression, we have assumed the field to have reached its desired final amplitude $\alpha_f$. With this replacement, neglecting quantum fluctuations of the resonator and taking $d=0$, \cref{eqn:Longitudinal_hamiltonian} is transformed to
\begin{equation}\label{eqn:Classical_H}
    \begin{split}
        \Tilde{H} = \frac{1}{2}\tilde n_t^2 - \{1-2 \sin^2[\varepsilon_p \cos(\Tilde{\omega}_d \Tilde{t})] \}\cos\tilde\varphi_t,
    \end{split}
\end{equation}
corresponding to a parametrically driven pendulum~\cite{mclaughlin1981,Hastings1993}.

In \cref{fig:chaos}(a, c-h), we present Poincaré sections obtained by plotting the phase space coordinates at integer multiples of the drive period $\tilde T = 2 \pi / \Tilde{\omega}_d$, as obtained from solving Hamilton’s equations for the dynamics governed by \cref{eqn:Classical_H}. Panel (a) is obtained in the absence of the parametric drive, $\varepsilon_p = 0$.  In that case, all trajectories on the Poincare section describe regular motion of the pendulum. Below the energy $\Tilde{H} < 1$, the trajectories correspond to oscillatory motion around the minimum of energy. For a deep enough transmon, i.e., small impedance $z$, these are in direct correspondence with the transmon bound states \cite{dumas2024unified}. On the other hand, above the energy $\Tilde{H}>1$ the trajectories correspond to full swings of the pendulum around its pivot point, corresponding to phase slips in the quantum model~\cite{Reminiscence_chaos}. At the energy $\Tilde{H}=1$ a single trajectory, the separatrix, connecting the points $(\tilde \varphi_t, \tilde n_t) = (\pm\pi,0)$, defines a boundary between these two distinct types of motion, see the green line on \cref{fig:chaos}(a). 

Panels (c), (e) and (g) show the Poincaré section for increasing drive amplitudes corresponding to $|\alpha_f|^2 \sim 1$, 9 and 49 photons, respectively. At these nonzero drive amplitudes, the aforementioned structure of the Poincare surface is altered, and two new types of motions can emerge, (i) chaos and (ii) subharmonic resonances, see Ref.~\cite{Reminiscence_chaos,dumas2024unified} for further details in the context of the driven transmon. The first phenomenon results from the fragility of the separatrix against integrability breaking perturbations. In short, the branches of outgoing and ingoing motion meet at a single point $(\varphi, n) = (\pi,0)$, and any arbitrarily weak integrability breaking perturbation splits this arrangement which marks the onset of an emergent chaotic region -- aka separatrix chaos, see chapter 6 of Ref.~\cite{Zaslavsky2005}. In addition to chaotic regions, the other type of motion that emerges in the presence of the drive is subharmonic resonances, which are characterized by the appearance of islands of regular motion within the chaotic region. These are known as $(n\!:\!m)$ nonlinear resonances and are characterized by the ratio $n:m$ between the drive-amplitude-dependent frequency of the pendulum and the drive frequency, with $n$ and $m$ integers~\cite{Zaslavsky2005}. With increasing drive amplitude, the chaotic region grows and the relative importance of $(n\!:\!m)$ resonances increasing, resulting in the central regular region shrinking. As observed in panels (d), (f), and (h), the same phenomenology is observed for the dispersive readout, see Ref.~\cite{Reminiscence_chaos,dumas2024unified} for detailed discussions. Crucially, however, for that readout the chaotic region rapidly engulfs much of phase space, dramatically impacting the regular region at the center of the phase space. This is a direct consequence of the different form of the drive in the two readouts. 

To understand how this phenomenology affects the qubit states, we follow Refs.~\cite{Reminiscence_chaos,dumas2024unified} and apply the Bohr-Sommerfeld quantization rule to identify the set of initial conditions (i.e., energies) corresponding to the ground and first excited states of the transmon. This rule identifies quantized orbits for oscillatory motion as the classical orbits that enclose an area given by $A_{j_t} = 2 \pi \hbar_\mathrm{eff} (j_t + 1/2)$~\cite{Messiah2020Quantum}, where $j_t = 0, 1, 2, \ldots$ denotes the Bohr-Sommerfeld quantum number for the transmon state, and $\hbar_\mathrm{eff} = z$ is the effective Planck constant~\cite{Reminiscence_chaos}. The total number of Bohr-Sommerfeld states within a given area $A$ can be calculated as $\lfloor A / 2 \pi \hbar_\mathrm{eff} \rfloor$. The orbits corresponding to the transmon’s ground and first excited states are depicted in \cref{fig:chaos}(a, c-h) by blue and red lines, respectively. These are regular orbits, enclosing areas $A_0 = \pi \hbar_\mathrm{eff}$ for the ground state and $A_1 = 3\pi \hbar_\mathrm{eff}$ for the first excited state. In panel (h) corresponding to a drive of 49 photons under dispersive readout, the size of the chaotic layer is such that the regular region is too small to support the regular orbits corresponding to the logical transmon states. In Refs.~\cite{dumas2024unified,Reminiscence_chaos}, this was identified as a signature of transmon ionization and linked to a critical photon number. The observation that the qubit orbits are no longer present in panel (h) for the dispersive readout is in stark contrast to the longitudinal readout where, for the same photon number, those orbits remain; compare panels (g) and (h). 

To compare more precisely the interplay of chaos with the computational subspace in the two readout schemes, we quantify the deformation of trajectories in phase space in the presence of a drive by computing for an initial condition $(\varphi(0),n(0))$ the deviation 
\begin{equation}\label{eqn:chaos_trajectory_deviation}
    \begin{split}
        d_{\alpha} = \sum_{t = 0}^{t_{\rm{max}}} \sqrt{\left[\varphi_0(t) - \varphi_\alpha(t)\right]^2 + \left[n_0(t) - n_\alpha(t)\right]^2}, 
    \end{split}
\end{equation}
where the subscripts $0$ and $\alpha$ indicate the classical trajectories of the undriven and driven case with amplitude $\alpha$, respectively, and $t_{\rm{max}}$ is the total evolution time. \Cref{fig:chaos}(b) presents this quantity as a function of photon number for both dispersive (purple) and longitudinal (blue) readouts. Here, the deviation is averaged over initial conditions randomly sampled from the phase space region spanning $[\pi \hbar_{\rm{eff}}, 4 \pi \hbar_{\rm{eff}}]$. This region corresponds to the area of phase space that contains the Bohr-Sommerfeld orbits of the transmon logical subspace. As is made clear by this figure and can be intuitively understood from the Pointcaré sections of panels (c-g) and (d-h), the dispersive readout is affected to a much larger extent and at much lower photon number than the longitudinal readout. In fact, a sudden jump in the average distance is observed for the dispersive readout at approximately 10 photons, while the longitudinal readout remains stable up to approximately 60 photons, and then only a small average distance is observed. In short, a parametrically driven transmon, corresponding to the longitudinal readout, is more robust against the emergence of chaos than a charge-driven transmon, corresponding to the dispersive readout. Interestingly, this conclusion applies to many scenarios involving parametric drives beyond longitudinal readout. 

\section{Conclusion}

We have studied the impact of multi-photon transitions in the transmon longitudinal readout \cite{Didier_longitudinal,Mathieu_thesis}. We find the threshold for ionization to be higher than for the dispersive readout. Notably, this threshold can be further increased by increasing the detuning between the qubit and the resonator, and that this can be done without sacrificing the amplitude of the dispersive shift. We have also found the longitudinal readout to be robust against disorder in the circuit. Those conclusions, obtained from a combination of numerical simulations, branch analysis, and Schrieffer-Wolff transformation, are supported by a qualitative analysis of the classical model of the transmon-resonator system. This analysis also showing that the longitudinal readout, which maps to a parametrically driven nonlinear oscillator, is more robust against the onset of chaos than the dispersive readout. We hope that these conclusions will motivate experimental exploration of the longitudinal readout as an alternative to the dispersive readout. Finally, the same tools can be used to further explore other alternative approaches to qubit readout, such as cross-Kerr readout \cite{Remy_cross_kerr}, quarton-coupler-based cross-Kerr readout \cite{ye2024ultrafast, Quarton_ye_2021}, and flux qubit readout \cite{Wang_flux_qubit_readout}. 

\section{Acknowledgments}
The authors are grateful to Othmane Benhayoune-Khadraoui, Benjamin D'Anjou, Crist\'obal Lled\'o, and Benjamin Groleau-Par\'e for helpful discussions. Support is acknowledged from NSERC, the Canada First Research Excellence Fund, the Minist\`ere de l’\'Economie et de l’Innovation du Qu\'ebec, Fonds de recherche du Qu\'ebec - Nature et technologies, and the U.S. Army Research Office Grant No. W911NF-22-S-0006. 

\appendix

\section{Circuit quantization} \label{app_sec:circuit_quantization}

In this section, we quantize the circuit of \cref{fig:longitudinal_circuit} following the standard approach of Ref.~\cite{Vool_Devoret}, see Ref.~\cite{Remy_cross_kerr} and supplementary information of Ref.~\cite{Lu_parity_protected} for a similar analysis of the circuit. Using this approach, we find the kinetic energy of the circuit
\begin{align}
    K = \frac{C_1}{2} \Dot{\phi}_1^2 + \frac{C_2}{2} \Dot{\phi}_2^2 + \frac{C_r}{2}(\Dot{\phi}_1 - \Dot{\phi}_2)^2, \label{eqn:cq_kinetic_pre}
\end{align}
and potential energy
\begin{align}
    U = E_{J1} \cos{\phi_1} + E_{J2} \cos{\phi_2} - \frac{1}{2 L_r}\left( \phi_1 - \phi_2 + \Phi_{\text{ext}} \right)^2. \label{eqn:cq_potential_pre}
\end{align}
Following the discussion in the main text, it is useful to define the two modes
\begin{align}
    \phi_t = \frac{\phi_1 + \phi_2 + \Phi_{\text{ext}}}{2}, \quad\quad \phi_r = \frac{\phi_1 - \phi_2 + \Phi_{\text{ext}}}{2}.
\end{align}
Substituting the above into \cref{eqn:cq_kinetic_pre} and \cref{eqn:cq_potential_pre}, we obtain the Lagrangian
\begin{align}
    \mathcal{L} &= \frac{C_1}{2} \left( \Dot{\phi}_t + \Dot{\phi}_r - \Dot{\Phi}_{\text{ext}} \right)^2 + \frac{C_2}{2} \left( \Dot{\phi}_t - \Dot{\phi}_r \right)^2 \nonumber \\ \nonumber &+ \frac{C_r}{2} \left( 2 \Dot{\phi}_r - \Dot{\Phi}_{\text{ext}} \right)^2 + E_{J1} \cos{(\phi_t + \phi_r - \Phi_{\text{ext}})} \\ &+ E_{J2} \cos{(\phi_t - \phi_r)} - \frac{2}{L_r} \phi_r^2.
\end{align}

The conjugate variables $q_t$ and $q_r$ of the above two mode flux are
\begin{align}
    q_t &= \frac{\partial \mathcal{L}}{\partial \dot{\phi}_t} = C_1 (\Dot{\phi}_t + \Dot{\phi}_r) + C_2 (\Dot{\phi}_t - \Dot{\phi}_r), \\
    q_r &= \frac{\partial \mathcal{L}}{\partial \dot{\phi}_r} = C_1 (\Dot{\phi}_t + \Dot{\phi}_r) - C_2 (\Dot{\phi}_t - \Dot{\phi}_r) + 4 C_r \Dot{\phi}_r,
\end{align}
corresponding to the capacitance matrix
\begin{align}
    \mathbf{C} = \begin{pmatrix}
    C_1 + C_2 & C_1 - C_2 \\
    C_1 - C_2 & C_1 + C_2 + 4 C_r
    \end{pmatrix}.
\end{align}
Assuming $C_1 = C_2 = C$, the inverted capacitance matrix is 
\begin{align}
    \mathbf{C}^{-1} = \begin{pmatrix}
    \frac{1}{2C} & 0 \\
    0 & \frac{1}{2C + 4C_r}
    \end{pmatrix}.
\end{align}
In terms of the above quantities, the Hamiltonian of the system takes the general form $H = \frac{1}{2} \mathbf{q}^T \mathbf{C}^{-1} \mathbf{q} - U$ where $\mathbf{q} = (q_t, q_r)$, resulting in, 
\begin{align}
    H &= \frac{q_t^2}{4 C} + \frac{q_r^2}{4C + 8C_r} + \frac{2}{L_r} \phi_r^2 \nonumber \\ &- E_{J1} \cos{(\phi_t + \phi_r - \Phi_{\text{ext}})} - E_{J2} \cos{(\phi_t - \phi_r)}.
\end{align}
Promoting the two pairs of conjugate variables to operators, and introducing the charge number operator $\hat{n}_{t,r} = \hat{q}_{t,r} / 2 e$, the Hamiltonian can be rewritten as,

\begin{align}
    \hat{H} &= \frac{4 e^2 \hat{n}_t^2}{4 C} + \frac{4 e^2 \hat{n}_r^2}{4 C + 8 C_r} + \frac{2}{L} \hat{\phi}_r \nonumber \\ &- E_{J1} \cos{(\hat{\phi}_t + \hat{\phi}_r - \Phi_{\text{ext}})} - E_{J2} \cos{(\hat{\phi}_t - \hat{\phi}_r)}.
\end{align}
Defining the transmon's charging energy $E_{C,t} = e^2 / 4 C$, resonator's charging energy $E_{C,r} = e^2 / (4 C + 8 C_r)$, and the resonator's inductive energy as $E_{L_r} = \Phi_0^2 / \pi^2 L_r$, the Hamiltonian takes the form
\begin{align}
    \hat{H} &= 4 E_{C,t} \hat{n}_t^2 + 4 E_{C,r} \hat{n}_r^2 + \frac{E_{L_r}}{2} \hat{\varphi}_r^2 \nonumber \\ &- E_{J1} \cos{(\hat{\varphi}_t + \hat{\varphi}_r - \varphi_{\text{ext}})} - E_{J2} \cos{(\hat{\varphi}_t - \hat{\varphi}_r)} \\
    &= \omega_r \hat{a}^\dagger \hat{a} + 4 E_{C,t} \hat{n}_t^2 - E_{J1} \cos{(\hat{\varphi}_t + \hat{\varphi}_r - \varphi_{\text{ext}})} \nonumber \\ &- E_{J2} \cos{(\hat{\varphi}_t - \hat{\varphi}_r)}, \label{eqn:cq_full_H}
\end{align}
where in the second line we have introduced the resonator's creation (annihilation) operators $\hat{a}^\dagger (\hat{a})$, and $\omega_r = \sqrt{8 E_{C,r} E_{L_r}}$ is the resonator frequency. We have also introduced the operators $\hat{\varphi}_{t,r} = (2 \pi / \Phi_0) \hat{\phi}_{t,r}$ and $\varphi_{\text{ext}} = (2 \pi / \Phi_0) \phi_{\text{ext}}$.

Using trigonometric identities, the last two terms of \cref{eqn:cq_full_H} can be expressed 
\begin{equation}\label{eqn:cq_coscos_expanded}
    \begin{split}
        E_{J1} & \cos{(\hat{\varphi}_t  + \hat{\varphi}_r - \varphi_{\text{ext}})} + E_{J2} \cos{(\hat{\varphi}_t - \hat{\varphi}_r)} 
        = \\ 
        &\left( E_{J1} \cos{\varphi_{\text{ext}}} + E_{J2} \right) \cos{\hat{\varphi}_t} \cos{\hat{\varphi}_r} \\ &+ \left( E_{J2} - E_{J1}  \cos{\varphi_{\text{ext}}} \right) \sin{\hat{\varphi}_t} \sin{\hat{\varphi}_r} \\ 
        &+ E_{J1} \sin{\varphi_{\text{ext}}} \cos{\hat{\varphi}_t} \left( \sin{\hat{\varphi}_r} - \cos{\hat{\varphi}_r} \right). 
    \end{split}
\end{equation}
Taking the external flux to be $\varphi_{\text{ext}} = 0$, the above expression can be reduced to
\begin{align}
    E_J \cos{\hat{\varphi}_t} \cos{\hat{\varphi}_r} + d E_J \sin{\hat{\varphi}_t} \sin{\hat{\varphi}_r},
\end{align}
where $E_J = E_{J1} + E_{J2}$ and $d = (E_{J2} - E_{J1}) / E_J$ is the asymmetry between the Josephson junction energies. Using these results, the Hamiltonian in \cref{eqn:cq_full_H} can be written as
\begin{align}
    \hat{H} &= \omega_r \hat{a}^\dagger \hat{a} + 4 E_C \hat{n}_t^2 - E_J \cos{\hat{\varphi}_t} \cos{\hat{\varphi}_r} \nonumber \\ &- d \sin{\hat{\varphi}_t} \sin{\hat{\varphi}_r} \\ 
    &= \omega_r \hat{a}^\dagger \hat{a} + 4 E_C \hat{n}_t^2 - E_J \cos{\hat{\varphi}_t} \nonumber \\ &+ 2 E_J \cos{\hat{\varphi}_t} \sin^2{(\hat{\varphi}_r / 2)} - d \sin{\hat{\varphi}_t} \sin{\hat{\varphi}_r}, \label{eqn:cq_final_H}
\end{align}
where in the second line we have again used a trigonometric identity to separate the bare transmon Hamiltonian $\hat{H}_t = 4 E_C \hat{n}_t^2 - E_J \cos{\hat{\varphi}_t}$ from the transmon-resonator interaction which can be read in the last line.

When including the gate charge $n_g$, we obtain the Hamiltonian of \cref{eqn:Longitudinal_hamiltonian} of the main text where we have also set $E_{C,t} = E_C$ to simplify the notation. We note that to drive the transmon (resonator) mode of this multi-mode circuit, the nodes labeled 1 and 2 in \cref{fig:longitudinal_circuit} need to be driven symmetrically (anti-symmetrically). However, if there is sufficient detuning between the two modes ($\Delta = \omega_q - \omega_r$ is large), it may be possible to drive only one mode of the circuit in order to control both the transmon and the resonator modes, independently with minimal cross-talk. Once again, a large detuning is beneficial for this circuit.

To get a more intuitive understanding of the above Hamiltonian, \cref{eqn:cq_final_H} can be reduced to its two-level approximation form, which we use in the main text to relate to the dispersive readout Hamiltonian. To obtain this reduced Hamiltonian, we first approximate the transmon Hamiltonian $\hat{H}_t$ as $(\omega_q / 2) \hat{\sigma}_z$, and assume $d = 0$ for simplicity. Then, using the fact that the zero point fluctuation of the transmon and resonator is small, the interaction term $2 E_J \cos{\hat{\varphi}_t} \sin^2{(\hat{\varphi}_r / 2)}$ can be expanded to
\begin{equation}
    \begin{split}
        2 E_J & \cos{\hat{\varphi}_t} \sin^2{(\hat{\varphi}_r / 2)} \approx  \\ 
        &2 E_J \left( 1 - \frac{\zpft^2}{2} \left( \hb + \hbd \right)^2 \right) \frac{\zpfr^2}{4} \left( \ha + \had \right)^2,
    \end{split}
\end{equation}
Using the rotating wave approximation as well as ignoring all terms that are proportional to $\hbd \hb$ and $\had \ha$ (which only correspond to a frequency shift to the transmon and resonator, respectively) we are left with
\begin{align}
    \hat{H}_{\text{int}} \approx - \frac{\zpft^2 \zpfr^2}{2} \hbd \hb \had \ha. 
\end{align}
Using $\hbd \hb = (1/2)(1 + \hat{\sigma}_z)$, $\zpft = (2 E_C / E_J)^{1/4}$ and setting $\chi_z = - \sqrt{2 E_C E_J} \zpfr^2 / 2$, we can write the two-level approximated Hamiltonian as, 
\begin{align}
    \hat{H}_{\text{two-level}} \approx \frac{\omega_q}{2} \hat{\sigma}_z + \omega_r \had \ha + \chi_z \hat{\sigma}_z \had \ha,
\end{align}
where we've lumped the frequency shift of the transmon (resonator) resulting from the interaction into $\omega_q$ ($\omega_r$). Note that the above dispersive shift $\chi_z$ is not obtained with a perturbative transformation, and depends only on the static circuit parameters such as the transmon charging and Josephson energy, as well as the phase zero-point fluctuation of the resonator (provided that $\varphi_{\rm{ext}} = 0$). Given that the phase zero-point fluctuation of the resonator can be kept constant whilst changing the resonator frequency, this means that the dispersive shift $\chi_z$ is truly independent of the detuning $\Delta$ between the qubit and the readout resonator. The benefits of breaking this trade-off between the magnitude of the dispersive shift and the detuning are highlighted in the main text, as well as below.

\section{Pulse shape} \label{app_sec:pulse_shape}

\begin{figure}
    \centering
    \includegraphics[width=\linewidth]{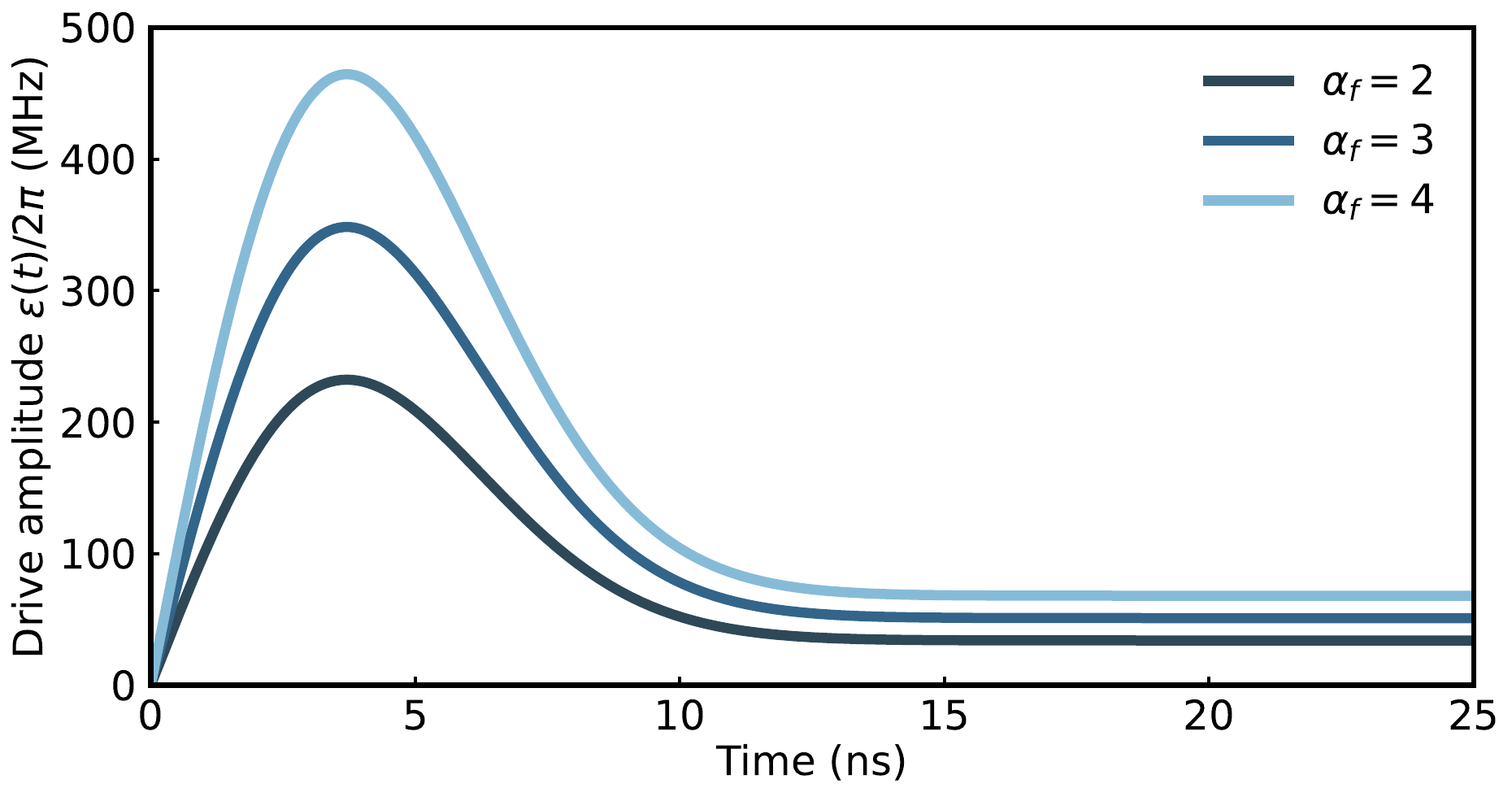}
    \caption{Pulses used for the readout simulations in \cref{fig:RO_sim}, shown up to $t = 25$ ns. Here $\tau = 5$ and $\kappa / 2\pi = 17$ MHz.}
    \label{fig:pulse_shape}
\end{figure}

To have a fast turn-on of the longitudinal interaction, we use a pulse shape $\epsilon(t)$ designed to rapidly fill the cavity with photons. Under the usual master equation accounting for damping of the resonator at the rate $\kappa$, the equation of motion for $\langle \ha \rangle$ is 
\begin{align}
    \langle \dot{\ha} \rangle = -i \omega_r \langle \ha \rangle - \frac{\kappa}{2} \langle \ha \rangle + \epsilon(t) \cos{(\omega_d t)}.
\end{align}
Defining $\langle \ha \rangle = \alpha(t)$, we can factor the solution of the above equation to slow and fast oscillating terms such that $\alpha(t) = \alpha_p(t) e^{-i \omega_r t}$. Substituting this into the above equation, and applying the rotating wave approximation, we get that 
\begin{align}
    \dot{\alpha}_p(t) = -\frac{\kappa}{2} \alpha_f(t) + \frac{\epsilon(t)}{2}. \label{eqn:app_alpha_p}
\end{align}
Using this expression, we choose a desired $\alpha_p(t)$ and obtain the corresponding form of $\epsilon(t)$. Here, we take $\alpha_p(t) = \alpha_f (1 - \exp{(-(t/\tau)^2)})$
with $\alpha_f$ the target coherent state amplitude. From \cref{eqn:app_alpha_p} we can then express the pulse as
\begin{align}
    \epsilon(t) = \frac{4 t \alpha_f}{\tau^2} e^{-(t / \tau)^2} + \kappa \alpha_f (1 - e^{-(t/\tau)^2}).
\end{align}
The above pulse is designed to quickly displace the resonator to a coherent state with an amplitude $\alpha_f$, and maintain it by accounting for dissipation. The speed at which the displacement is made is controlled by adjusting $\tau$. The pulse shapes used in \cref{fig:RO_sim} are shown in \cref{fig:pulse_shape}. For $\alpha_f = 4$, the maximum drive amplitude is $\varepsilon_{\rm{max}} / 2\pi \approx 465$ MHz, and is reached in approximately $5$ ns.

It is important to point out that the phase of the drive must be constant in order to activate the longitudinal interaction most effectively. Apart from experimental limitations, there are no restrictions on what the pulse shape may be, and it could be interesting to consider if more sophisticated pulses could further improve the performance, just like in dispersive readout \cite{Walter_disp_readout, gautier2024optimal}.

\section{Critical photon number identification} \label{app_sec:ncrit_classification}

We identify ionization critical photon number of the ground (excited) state of the transmon from the branch analysis as the number of photons in the resonator that is required for the transmon population to exceed $N_t = 2$ ($N_t = 3$), for at least three photons in a row; see Ref.~\cite{dumas2024unified} for details. The last condition is to avoid identifying small resonances that can bring the population over the threshold at a single resonator photon number. See, for example, the small population jump due to a small resonance between states $1_t$ and $5_t$ in \cref{fig:branch_analysis_1}(c).

\section{Schrieffer-Wolff transformation} \label{app_sec:SW}

In this section, we derive the Schrieffer-Wolff transformation for the longitudinal readout Hamiltonian \cref{eqn:Longitudinal_hamiltonian}, assuming $d = 0$ and in the absence of a  drive. We express this Hamiltonian as
\begin{align}
    \h{H} = \h{H}_0^\prime + \h{V}^\prime,
\end{align}
with 
\begin{align}
    \h{H}_0^\prime &= \omega_r \had \ha + \sum_{j_t} \omega_{j_t} \ket{j_t} \bra{j_t},
\end{align}
and
\begin{equation}
\h{V}^\prime   = 2 E_J \cos{\hvt} \sin^2{(\hvr / 2)} 
\end{equation}
where $\ket {j_t}$ are the transmon eigenstates and $\omega_{j_t}$ are the corresponding energies. Note that $\hat{V}^\prime$ has diagonal components in the basis in which we have expressed $\hat{H}_0^{\prime}$. We separate out this diagonal contribution, from which we define $\hat{H}_0$ in the main text. The off-diagonal part are then, by definition, $\hat{V}$. Using the trigonometric identity $2 \sin^2\theta/2 = 1 - \cos \theta $, we can express $\hat{V}^\prime$ as a sum of displacements operators $\hat{D}(\alpha)$  
\begin{equation}
    \begin{split}
        \cos{\hvr} 
        &= \frac{1}{2} \left( e^{i \zpfr (\ha + \had)} + e^{-i \zpfr (\ha + \had)} \right) \\ 
        &= \frac{1}{2} \left[ D(i \zpfr) + D(-i \zpfr) \right].
    \end{split}
\end{equation}

The matrix elements of the displacement operator in the Fock basis are well-known and take the form \cite{Cahill_Glauber} 
\begin{align}
    \langle n_r \vert D(\alpha) \vert m_r \rangle = \sqrt{\frac{m_r !}{n_r !}} \alpha^{n_r - m_r} e^{-|\alpha|^2 / 2} L_{m_r}^{n_r - m_r}(|\alpha|^2),
\end{align}
where $n_r \geq m_r$, and $L_a^b$ is the generalized Laguerre polynomial (with $a$ a nonnegative integer and $b$ a real number), we can write the interaction term as
\begin{widetext}
\begin{align}
    \h{V}^\prime &= E_J \sum_{i_t, j_t} \langle i_t \vert \cos{\hvt} \vert j_t \rangle \ket{i_t} \bra{j_t} \\ 
    &-\frac{E_J}{2} e^{-\zpfr^2 / 2} \sum_{i_t, j_t} \sum_{n_r, m_r} \langle i_t \vert \cos{\hvt} \vert j_t \rangle \sqrt{\frac{m_r !}{n_r !}} (i \zpfr)^{n_r - m_r} L_{m_r}^{n_r-m_r}(\zpfr^2) \left[ 1 + (-1)^{n_r - m_r} \right] \ket{i_t, n_r} \bra{j_t, m_r}. \label{app_eqn:resonator_selection_rules}
\end{align}
The total Hamiltonian can then be written as in the main text
\begin{equation}\label{eq:app_H_with_offdiag}
    \hat{H} = \hat{H}_0^\prime + \hat{V}^\prime  =\hat{H}_0 + \h{V},
\end{equation}
with
\begin{equation}\label{app_eqn:H0_diag} 
    \begin{split}
        \hat{H}_0 &= \omega_r \had \ha + \sum_{j_t} \left( \omega_{j_t} + E_J \langle j_t \vert \cos{\hat{\varphi}_t} \vert j \rangle \right) \vert j_t \rangle \langle j_t \vert - E_J e^{-\zpfr^2 / 2} \sum_{j_t} \sum_{m_r} \langle j_t \vert \cos{\hat{\varphi}_t} \vert j_t \rangle L_{m_r}(\zpfr^2)  \vert j_t, m_r \rangle \langle j_t, m_r \vert \\
    &= \sum_{j_t} \left[ \omega_r + \chi_{z, j_t}(\had \ha) \ket{j_t} \bra{j_t} \right] \had \ha + \sum_{j_t} \left( \omega_{j_t} + \Lambda_{j_t} \right) \ket{j_t} \bra{j_t},  
    \end{split}
\end{equation}
where 

\begin{align}
    \had \ha \chi_{z, j_t}(\had \ha) = -E_J e^{-\zpfr^2/2} L_{\had \ha}(\zpfr^2) \sum_{j_t} \langle j_t \vert \cos{\hat{\varphi}_t} \vert j_t \rangle 
\end{align}
is the transmon dependent shift to the resonator frequency and 
\begin{align}
    \Lambda_{j_t} = E_J \langle j_t \vert \cos{\hat{\varphi}_t} \vert j_t \rangle,
\end{align}
is a resonator-independent shift of the transmon's frequency. Moreover, the off-diagonal part of the interaction term can then be written as 
\begin{equation}
    \begin{split}
        \h{V} &= E_J \sum_{i_t \neq j_t} \langle i_t \vert \cos{\hat{\varphi}_t} \vert j_t \rangle \vert i_t \rangle \langle j_t \vert \\ 
        &- \frac{E_J}{2} e^{-\zpfr^2 / 2} \sum_{(i_t, n_r) \neq (j_t, m_r)} \langle i_t \vert \cos{\hat{\varphi}_t} \vert j_t \rangle \sqrt{\frac{m_r!}{n_r!}} (i \zpfr)^{n_r - m_r} L_{m_r}^{n_r - m_r}(\zpfr^2) \left[ 1 + (-1)^{n_r - m_r} \right] \vert i_t, n_r \rangle \langle j_t, m_r \vert. 
    \end{split}
\end{equation}

Having separated out the diagonal and off-diagonal contributions, we now perform a Schrieffer-Wolff transformation on \cref{eq:app_H_with_offdiag} to eliminate the off-diagonal contribution to second order to obtain the effective Hamiltonian 
\begin{align}
    e^{\hat{S}} \hat{H} e^{-\hat{S}} \approx \hat{H}_0 + \hat{H}^{(2)} + \dots, 
\end{align}
with $\hat{S}$ the Schrieffer-Wolff transformation generator, and 
\begin{align}
    \hat{H}^{(2)} = \sum_{i_t, j_t, k_t} \sum_{n_r, m_r, l_r} \frac{1}{2} \eta_{i_t, k_t, n_r, l_r} \eta_{k_t, j_t, l_r, m_r} \left( \frac{1}{\omega_{i_t, k_t} + (n_r - l_r) \omega_r^\prime } + \frac{1}{\omega_{j_t, k_t} + (m_r - l_r) \omega_r^\prime} \right) \vert i_t, n_r \rangle \langle j_t, m_r \vert, \label{app_eqn:H2}
\end{align}
with $\omega_{a_t, b_t} = \omega_{a_t} - \omega_{b_t}$ where $\omega_{a_t}$ and $\omega_{b_t}$ are the ac-Stark shifted transmon frequencies, $\omega_r^\prime$ the resonator frequency shifted by $\chi_{z, j_t}(\had \ha)$, and the matrix elements $\eta_{a_t, b_t, c_r, d_r}$ given by
\begin{equation}
    \begin{split}
        \eta_{a_t, b_t, c_r, d_r} &= E_J \langle a_t \vert \cos{\hat{\varphi}_t} \vert b_t \rangle \delta_{c_r, d_r} \\
    &- \frac{E_J}{2} e^{-\zpfr^2 / 2} \langle a_t \vert \cos{\hat{\varphi}_t} \vert b_t \rangle \sqrt{\frac{d_r !}{c_r !}} (i \zpfr)^{c_r - d_r} L_{d_r}^{c_r - d_r}(\zpfr^2) \left[ 1 + (-1)^{c_r - d_r} \right].
    \end{split}
\end{equation}
\end{widetext}
It is worth noting that the selection rules for both the transmon and the resonator are now apparent. As mentioned in the main text, the cosine function is even and, therefore, only states with the same parity can be connected with matrix elements $\langle i_t \vert \cos{\hvt} \vert j_t \rangle$. We observe similar selection rules for the resonator. This can be seen in \cref{app_eqn:resonator_selection_rules} where the term $[ 1 + (-1)^{n_r - m_r}]$ only connects resonator states if $n_r - m_r$ is even. These selection rules are beneficial for the readout as they prevent unwanted transitions between the computational states, and reduce the numbers of pathways for transitions to higher states to occur. 

Furthermore, the diagonal Hamiltonian of \cref{app_eqn:H0_diag} includes the Lamb shift, as well as the cross-Kerr coupling that is used for the readout. As mentioned in the main text, an important point is that the cross-Kerr coupling does not depend on the transmon-resonator detuning. Moreover, the off diagonal terms that lead to ionization are in \cref{app_eqn:H2}. Crucially, unlike the cross-Kerr coupling, however, these terms \textit{do} depend on the detuning between the transmon and the resonator. Therefore, increasing the detuning can then suppress the terms that lead to ionization. 

\section{Readout simulations} \label{app_sec:stochastic_readout_sims}

To model the readout and obtain an accurate estimation of the assignment error, we numerically model a single-shot readout measurement with heterodyne detection using QuTip's stochastic Schr\"odinger equation solver \cite{Qutip_1, Qutip_2}. The stochastic Schr\"odinger equation we are solving is,
\begin{align}
    d \psi(t) &= -i \h{H} \psi(t) dt \nonumber \\ 
    &- \left( \frac{\kappa}{2} \had \ha - \frac{\kappa}{4} [ \langle \h{x} \rangle + i \langle \h{p} \rangle ] \ha + \frac{\kappa}{16} [\langle \h{x} \rangle^2 + \langle \h{p} \rangle^2 ] \right) \psi(t) \nonumber \\
    &+ \sqrt{\frac{\kappa}{2}} \left( \ha - \frac{\langle \h{x} \rangle}{2} \right) \psi(t) \: dW_x \nonumber \\
    &- \sqrt{\frac{\kappa}{2}} \left( i \ha + \frac{\langle \h{p} \rangle}{2} \right) \psi(t) \: dW_p. \label{eqn:SSE}
\end{align}
Here, $\h{H}$ is the Hamiltonian from \cref{eqn:Longitudinal_Hamiltonian_original}, $\kappa$ is the resonator decay rate, and $\h{x} = \ha + \had$, and $\h{p} = i (\had - \ha)$ are the measured quadratures. Both measured quadratures have independent Wiener increments $d W_i$ which satisfy $\mathbb{E}[dW_i] = 0$ and $\mathbb{E}[dW_i dW_j] = \delta_{ij}dt$ where $i,j \in \{ x, p \}$. For each trajectory, we demodulate the signal with $e^{i \omega_d t}$, and integrate the signal from both the $x$ and $p$ quadratures using the real and imaginary part of the optimal weighting function $\langle a_e(t) \rangle - \langle a_g(t) \rangle$, respectively~\cite{Bultink2018}. 

\begin{figure}[h]
    \centering
    \includegraphics[width=\linewidth]{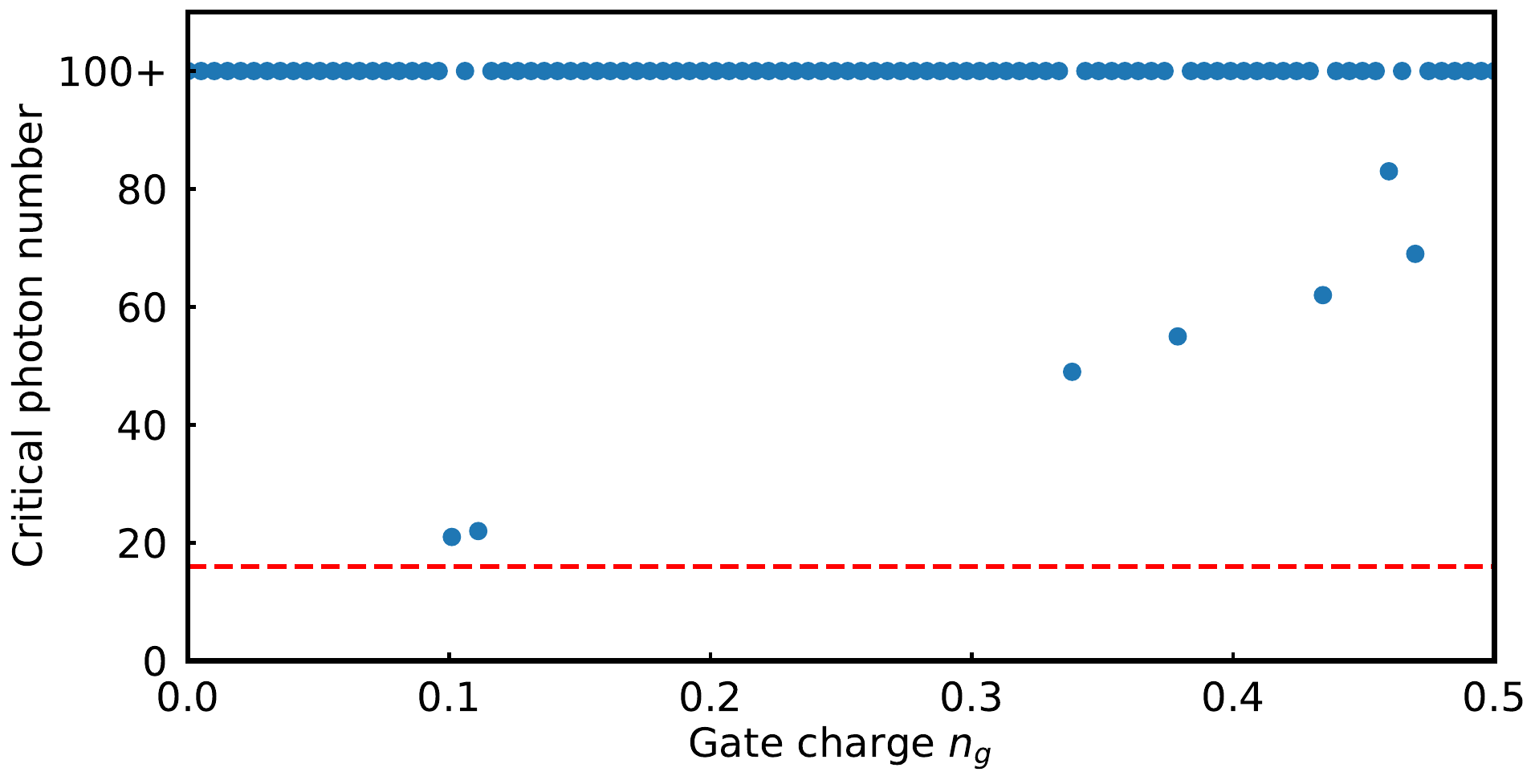}
    \caption{Critical photon number for the parameters used in the stochastic readout simulations for \cref{fig:RO_sim} as a function of varying gate charge $n_g$. Red line indicates the maximum number of photons $n_{\rm{max}}$ that we put in the resonator for the readout simulations. For all gate charges scanned, the critical photon number remains above $n_{\rm{max}}$.}
    \label{app_fig:readout_ncrit_gc}
\end{figure}

By simulating multiple trajectories, with the qubit initialized in the ground or the excited state, we obtain two cluster of points in the IQ plane as a function of time. At each time step $t$, we find the optimal threshold line that separates the two clusters of points which results in the minimal assignment error $\varepsilon$(t) computed as,  
\begin{align}
    \varepsilon(t) = \frac{P(e|g)(t) + P(g|e)(t)}{2},
\end{align}
where $P(x|y)(t)$ is the conditional probability to assign the readout outcome as $x$ when starting in the $y$th state. In our simulations, we use a step size of $\pi / \omega_d$, with nsubsteps set to $6000$ for a target coherent state amplitude of $\alpha_f = 2$, and $10000$ for $\alpha_f = 3$, and $4$. We simulated $81920$, $26624$, and $27648$ trajectories for the target coherent state amplitude $\alpha_f = 2$, $3$, and $4$, respectively. For the stochastic readout simulations we do not average over gate charges. However, as indicated in \cref{app_fig:readout_ncrit_gc}, we find that the critical photon number does not dip below $n_{\rm{max}} = 16$, the maximum number of photons we put in the resonator in our simulations.

Computing the assignment fidelity in the above way accounts for any non-Gaussian effects in the phase space that may be overlooked when using analytical expressions that assume the resonator fields as coherent states. However, given the relatively low self-Kerr on the resonator ($\approx 340$ kHz), we find that the non-Gaussian effects on the resonator fields are minimal at the photon numbers used in our readout simulations. We can then approximate the resonator states to be coherent states and compute the optimal signal-to-noise ratio with
\begin{align}
    \mathrm{SNR}(t_\mathrm{m}) = \sqrt{2 \kappa \eta \int_0^{t_\mathrm{m}} \vert \alpha_{e}(t) - \alpha_{g}(t) \vert^2 \: dt},
\end{align}
where $\eta$ is the measurement efficiency, $t_m$ is the integration time, and $\alpha_{g}(t)$ ($\alpha_{e}(t)$) is the resonator mean field when the qubit is prepared in the ground (excited) state. We compute the resonator mean field using QuTiP's Monte-Carlo solver with $500$ trajectories. The assignment fidelity can then be computed with
\begin{align}
    \varepsilon(t_m) = \frac{1}{2} \mathrm{erfc}\left(\frac{\mathrm{SNR}(t_m)}{2 \sqrt{2}}\right),
\end{align}
where $\mathrm{erfc}$ is the complementary error function. The estimated assignment errors are shown in \cref{fig:RO_sim} with the measurement efficiency set to $\eta = 0.5$.

\bibliography{articles.bib}

\end{document}